\documentclass[a4paper,amsmath,amssymb]{jpconf}
\usepackage{graphicx}
\usepackage{bm}
\usepackage{amsmath, amsthm, amssymb}
\usepackage[ruled]{algorithm}
\usepackage{algpseudocode}
\usepackage{algpascal}
\usepackage{algc}

\newcommand\ASTART{\bigskip\noindent\begin{minipage}[b]{0.5\linewidth}}

\newcommand\AENDSKIP{\end{minipage}\bigskip}
\newcommand\AEND{\end{minipage}}

\newcommand{\V}[1]{\bm{#1} } %vector command
\newcommand{\Ave}[1]{\left\langle {#1} \right\rangle} %thermal average 
\newcommand{\mR}{\mathbb{R}}

\newcommand{\lb}{\left(}
\newcommand{\rb}{\right)}
\newcommand{\lbb}{\left\{}
\newcommand{\rbb}{\right\}}

\newcommand{\Req}[1]{eq.\ (\ref{eq:#1})}
\newcommand{\BReq}[1]{Eq.\ (\ref{eq:#1})}
\newcommand{\NReq}[1]{(\ref{eq:#1})}

\newcommand{\Rfig}[1]{Fig.\ \ref{fig:#1}}
\newcommand{\Rfigss}[2]{Figs.\ \ref{fig:#1}-\ref{fig:#2}}
\newcommand{\Lfig}[1]{\label{fig:#1}}
\newcommand{\Leq}[1]{\label{eq:#1}}

\newcommand{\Lsec}[1]{\label{sec:#1}}
\newcommand{\be}{\begin{eqnarray}}
\newcommand{\ee}{\end{eqnarray}}
\newcommand{\ba}{\begin{array}}
\newcommand{\ea}{\end{array}}

\newcommand{\subbe}{\begin{subequations}}
\newcommand{\subee}{\end{subequations}}
\newcommand{\mc}[1]{\mathcal{#1}}
\DeclareMathOperator*{\argmin}{arg\,min}

\newcommand{\lA}{\leftarrow}

\newcommand{\Lcode}[1]{\label{code:#1}}
\newcommand{\Rcode}[1]{Alg.\ \ref{code:#1}}

\begin{document}
\title{Sparse approximation problem: how rapid simulated annealing succeeds and fails}

\author{Tomoyuki Obuchi$^{1}$ and Yoshiyuki Kabashima}

\address{Interdisciplinary Graduate School of Science and Engineering, Tokyo Institute of Technology, Yokohama, Kanagawa 226-8502, Japan}

\ead{$^{1}$obuchi@sp.dis.titech.ac.jp}

%%%%%%%%%%%%%%%%%%%%%%%%%%%%%%%%%%%%%%%%%%%%%%%%%%%%%%
%%%%%%%%%%%%%%%%%%%%%%%%%%%%%%%%%%%%%%%%%%%%%%%%%%%%%%
%%%%%%%%%%%%%%%%%%%%%%%%%%%%%%%%%%%%%%%%%%%%%%%%%%%%%%
\begin{abstract}
Information processing techniques based on sparseness have been actively studied in several disciplines.
Among them, a mathematical framework to approximately express a given dataset by a combination of a small number of basis vectors of an overcomplete basis is termed the {\em sparse approximation}. In this paper, we apply simulated annealing, a metaheuristic algorithm for general optimization problems, to sparse approximation in the situation where the given data have a planted sparse representation and noise is present. The result in the noiseless case shows that our simulated annealing works well in a reasonable parameter region: the planted solution is found fairly rapidly. This is true even in the case where a common relaxation of the sparse approximation problem, the $\ell_1$-relaxation, is ineffective. On the other hand, when the dimensionality of the data is close to the number of non-zero components, another metastable state emerges, and our algorithm fails to find the planted solution. This phenomenon is associated with a first-order phase transition. In the case of very strong noise, it is no longer meaningful to search for the planted solution. In this situation, our algorithm determines a solution with close-to-minimum distortion  fairly quickly. 
\end{abstract}

%%%%%%%%%%%%%%%%%%%%%%%%%%%%%%%%%%%%%%%%%%%%%%%%%%%%%%
%%%%%%%%%%%%%%%%%%%%%%%%%%%%%%%%%%%%%%%%%%%%%%%%%%%%%%
%%%%%%%%%%%%%%%%%%%%%%%%%%%%%%%%%%%%%%%%%%%%%%%%%%%%%%
\section{Introduction}\label{sec:Introduction}
The success of compressed sensing~\cite{Donoho:06-1,Candes:05,Candes:06a,Candes:06b} has triggered interest in the utilization of sparseness in signal processing techniques~\cite{Donoho:09-1,Donoho:09-2,Kabashima:09,Ganguli:10,Rangan:10,Krzakala:12,Sakata:13,Nakanishi:15,Obuchi:15}. Sparseness is the property whereby data can be represented, on a proper basis, by some combination of a small number of non-zero components. This property is useful for practical applications such as data compression and data reconstruction from a small number of observations, the latter of which is simply compressed sensing. 

Usually, obtaining a sparse representation from a given dataset is formulated as an optimization problem. We refer to this as the sparse approximation problem~\cite{Natarajan:95,Temlyakov:98,Temlyakov:99,Tropp:04,Donoho:06-2}. This is sometimes recast in a probabilistic formulation by statistical physicists~\cite{Krzakala:12,Nakanishi:15,Obuchi:15} using Bayesian techniques or statistical mechanics. In this paper, we employ such a probabilistic formulation to search for an ``optimal'' solution with the minimum distortion between the given and reconstructed data. 

Unlike previous approaches, we do not use a message passing algorithm~\cite{Krzakala:12}. Instead, we use  the well-known ``simulated annealing'' (SA) heuristic. The motivation for using SA comes from our recent theoretical analysis of ``entropy''~\cite{Nakanishi:15,Obuchi:15}, which is the exponential rate of the number of combinations of non-zero components yielding a given level of distortion. Our analysis indicates that entropy exhibits some nice analytical properties, in contrast to other optimizations such as the $k$-satisfiability problem~\cite{Monasson:96,Krzakala:07}. This implies a simple structure of the ``phase space,'' the space of possible combinations of non-zero components, and hence SA is expected to work well. Based on this expectation, the actual performance of SA is reported through numerical experiments. 

%%%%%%%%%%%%%%%%%%%%%%%%%%%%%%%%%%%%%%%%%%%%%%%%%%%%%%
%%%%%%%%%%%%%%%%%%%%%%%%%%%%%%%%%%%%%%%%%%%%%%%%%%%%%%
%%%%%%%%%%%%%%%%%%%%%%%%%%%%%%%%%%%%%%%%%%%%%%%%%%%%%%
\section{Formulation and algorithm}

%%%%%%%%%%%%%%%%%%%%%%%%%%%%%%%%%%%%%%%%%%%%%%%%%%%%%%
%%%%%%%%%%%%%%%%%%%%%%%%%%%%%%%%%%%%%%%%%%%%%%%%%%%%%%
\subsection{Combinatorial optimization formulation}
Let us suppose a signal vector $\V{y}\in \mR^{M}$ is generated from an appropriate sparse representation or a planted solution, $\hat{\V{x}}$, through
\be
\V{y}=A\hat{\V{x}}+\V{\xi},
\Leq{y-def}
\ee
where $A=\{\V{a}_i \}_{i=1}^{N} \in \mR^{M\times N}$ is an overcomplete matrix with $M<N$  called a dictionary; $\V{\xi}\in \mR^{M}$ is a noise vector, each component of which is drawn from the zero-mean normal distribution with variance $\sigma_{\xi}^2$, $\mathcal{N}(0,\sigma_{\xi}^2)$. The appropriate representation $\hat{\V{x}}$ is ``sparse,'' and is assumed to be generated from the following Bernoulli--Gaussian distribution: 
\be
P(\hat{x}_i)=\hat{\rho}\frac{  e^{ -\frac{1}{2\sigma_x^2}\hat{x}_i^2 }  }{\sqrt{ 2\pi \sigma_x^2} }+(1-\hat{\rho})\delta(\hat{x}_i).
\ee
We also assume that each component of $A$ is independent and identically distributed from $\mc{N}(0,1/N)$. The aspect ratio of the matrix, $\alpha=M/N$, is assumed to lie within $(0, 1)$.

The sparse approximation problem is to approximate $\V{y}$ by a linear combination of a restricted number of column vectors of $A$. There are various formulations,  one of which is based on the following optimization:
\be
\V{x}^*=\argmin_{\V{x}}\lbb \mc{E}(\V{x}|\V{y},A) \rbb~{\rm subject~to}~||\V{x}||_0\leq N\rho,
\Leq{optimization}
\ee
where $||\V{x}||_k=(\sum_{i}|x_i|^k)^{1/k}$ denotes the $\ell_k$ norm and the $\ell_0$ norm is equal to the number of non-zero components of $\V{x}$; the parameter $\rho(<\alpha)$ controls the sparseness of $\V{x}$ (and is called the sparsity in this paper); and $\mc{E}$ denotes the distortion between $\V{y}$ and a reconstructed signal through a representation $\V{x}$:
\be
\mc{E}(\V{x}|\V{y},A)=\frac{1}{2}||\V{y}-A\V{x}||_2^2.
\Leq{epsilon(x)}
\ee

%%%%%%%%%%%%%%%%%%%%%%%%%%%%%%%%%%%%%%%%%%%%%%%%%%%%%%
%%%%%%%%%%%%%%%%%%%%%%%%%%%%%%%%%%%%%%%%%%%%%%%%%%%%%%
\subsection{Probabilistic formulation}
\BReq{optimization} is a commonly used formulation, but it has the limitation that it only provides the information of the minimum-distortion solution. To get a wider perspective, it is better to treat {\it all} possible combinations of the column vectors. For this, we use a probabilistic formulation. Suppose that a binary vector  $\V{c}=\lbb c_i=0,1\rbb_{i=1}^{N}$, which we call the sparse weight, represents the column vectors used to represent $\V{y}$: if $c_i=1$, the $i$th column of $A$, $\V{a}_i$,  is used; if $c_i = 0$, it is not. Once these columns have been  determined by $\V{c}$, the optimal coefficients of the chosen columns are evaluated by solving
\be
\V{x}(\V{c})=\argmin_{\V{x}}||\V{y}-A(\V{c}\circ \V{x}) ||_2^2,
\Leq{x(c)}
\ee
where $(\V{c}\circ \V{x})_i=c_i x_i$ represents the Hadamard product. The corresponding distortion is 
\be
\mc{E}(\V{c}|\V{y},A)=M\epsilon(\V{c}|\V{y},A)=\frac{1}{2}||\V{y}-A(\V{c}\circ \V{x}(\V{c}))||_2^2.
\Leq{epsilon(c)}
\ee
The components of $\V{x}(\V{c})$ for the zero components of $\V{c}$ are actually indefinite, and we set them to be zeros. The definite part of $\V{x}(\V{c})$, which we denote $\tilde{\V{x}}(\V{c})$, has the compact analytic form
\be
\tilde{\V{x}}(\V{c})=\lb \tilde{A}^{\rm T}(\V{c})\tilde{A}(\V{c}) \rb^{-1} \tilde{A}^{\rm T}(\V{c})\V{y},
\ee
where $\tilde{A}(\V{c})$ denotes the submatrix of  columns chosen by $\V{c}$. 

Let us regard $\mc{E}$ as an ``energy,'' and introduce an ``inverse temperature'' $\mu$. A Gibbs--Boltzmann distribution is thus defined as
\be
P(\V{c}|\mu;\V{y},A)=\frac{1}{G(\mu;\V{y},A)}\delta\lb \sum_{i}c_i-N\rho \rb e^{-\mu \mc{E}(\V{c}|\V{y},A)},
\Leq{P(c)}
\ee
where $G$ is the ``partition function''
\be
G(\mu;\V{y},A)=\sum_{\V{c}}\delta\lb \sum_{i}c_i-N\rho \rb e^{-\mu \mc{E}(\V{c}|\V{y},A)}.
\ee
Our strategy is to generate $\V{c}$ according to \Req{P(c)}. Changing $\mu$ allows us to sample different sparse solutions with different distortion values. 

This formulation provides several options to treat the sparse approximation problem. For example, sampling in $\mu<\infty$ produces solutions with distortion greater than the minimum.  These ``finite temperature'' solutions may be more suitable for capturing the planted solution $\hat{\V{x}}$ than $\V{x}^*$ in \Req{optimization} in the presence of noise $\sigma_{\xi}>0$, as suggested in \cite{Obuchi:15}. 

In this probabilistic formulation, the optimization  \NReq{optimization} is recovered as a sampling problem in the limit $\mu \to \infty$. We pursue this direction in this paper, and propose an algorithm to solve \Req{optimization}. The performance of our technique is examined in numerical experiments, and is compared with some known analytical results~\cite{Kabashima:09,Nakanishi:15,Obuchi:15}.

%%%%%%%%%%%%%%%%%%%%%%%%%%%%%%%%%%%%%%%%%%%%%%%%%%%%%%
%%%%%%%%%%%%%%%%%%%%%%%%%%%%%%%%%%%%%%%%%%%%%%%%%%%%%%
\subsection{Simulated annealing}\Lsec{SA}
Our algorithm is a variant of SA, which is a metaheuristic to find the global minimum of a cost function. The outline of the algorithm is as follows: starting from a random initial configuration of $\V{c}$ at very high temperature $T=1/\mu \gg 1$, the algorithm randomly updates the configuration $\V{c}\to \V{c}'$ in a Monte-Carlo (MC) manner, while gradually decreasing the temperature. Eventually, the temperature becomes very low, $T\approx 0$, and the configuration is no longer updated. This final configuration is expected to be very close (or identical) to the true solution, {\it i.e.}, $\V{x}^*\approx \V{c}\circ \V{x}(\V{c})$. 

The Metropolis criterion is adopted in our simulation: an MC move $\V{c}\to \V{c}'$ is judged to be accepted or not according to the probability
\be
p_{\rm accept}(\V{c}\to \V{c}')=\max(1,e^{-\mu\lb \mc{E}(\V{c}') -\mc{E}(\V{c})\rb } ).
\ee
For a fixed value of $\rho$, the configurations generated by the algorithm should always satisfy $\sum_{i}c_i=N\rho$. Given an initial configuration satisfying this condition, we generate trial moves $\V{c}\to \V{c}'$ by ``pair flipping'' two sparse weights, one equal to $0$ and the other equal to $1$. Namely, choosing an index $i$ of the sparse weight from ${\rm ONES}\equiv \{k|c_{k}=1 \}$ and another index $j$ from ${\rm ZEROS}\equiv \{k|c_{k}=0  \}$, we set $\V{c}'=\V{c}$, except for the counterpart of $(c_i,c_j)=(1,0)$, which is given as $(c'_i,c'_j)=(0,1)$.

The pseudo-code of our MC algorithm is given in \Rcode{MC}, and that of our SA procedure is presented in \Rcode{SA}.
%%%%%%%%%%%%%%%%%%%%%%%%%%%%%%%
\alglanguage{pseudocode}
\begin{algorithm}[htbp]
\caption{MC update with pair flipping}\Lcode{MC}
\begin{algorithmic}[1]
\Procedure{MCpf}{$\V{c},\mu,\V{y},A$}\Comment{MC routine  with pair flipping}
	\State ${\rm ONES} \lA \{k|c_k=1\},~{\rm ZEROS} \lA \{k|c_k=0\}$
	\State randomly choose $i$ from ONES and $j$ from ZEROS
	\State $\V{c}' \lA \V{c}$
	\State $(c'_i, c'_j) \lA (0,1)$
	\State $(\mc{E},\mc{E}')\lA (\mc{E}(\V{c}|\V{y},A),\mc{E}(\V{c}'|\V{y},A))$ \Comment{Calculate energy}
	\State $p_{\rm accept} \lA \max(1,e^{-\mu\lb \mc{E}' -\mc{E}\rb } )$
	\State generate a random number $r\in [ 0,1]$
	\If {$ r < p_{\rm accept} $}
		\State $\V{c} \lA \V{c}'$
	\EndIf       
	\State \Return $\V{c}$
\EndProcedure
\end{algorithmic}
\end{algorithm}
%%%%%%%%%%%%%%%%%%%%%%%%%%%%%%%
%%%%%%%%%%%%%%%%%%%%%%%%%%%%%%%%
%\begin{algorithm}[htbp]
%\caption{Energy function $\mc{E}$}\Lcode{energy}
%\begin{algorithmic}[1]
%\Procedure{$\mc{E}$}{$\V{c},\V{y},A$}\Comment{Routine of MC with pair flipping}
%	\State ${\rm ONES} \lA \{c_{k}|c_k=1\}$
%   	\State $\tilde{A} \lA A({\rm ONES},{\rm ONES})$ \Comment{Submatrix for chosen columns by $\V{c}$}	
%	\State $\tilde{\V{x}} \lA  (\tilde{A}^{\rm T}\tilde{A})^{-1}\tilde{A}^{\rm T} \V{y}$ 
%	\Comment{Optimal weights for chosen columns by $\V{c}$}
%	\State $\Delta \V{y} \lA \V{y}-\tilde{A}\cdot \tilde{\V{x}}$
%	\State $E=\frac{1}{2}\Delta\V{y}\cdot \Delta\V{y}$
%	\State \Return $E$
%\EndProcedure
%\end{algorithmic}
%\end{algorithm}
%%%%%%%%%%%%%%%%%%%%%%%%%%%%%%%%
%%%%%%%%%%%%%%%%%%%%%%%%%%%%%%%
\begin{algorithm}[htbp]
\caption{SA for sparse approximation problem}\Lcode{SA}
\begin{algorithmic}[1]
\Procedure{SA}{$\{\mu_a,\tau_a \}_{a=1}^{N_{\mu}},\rho,\V{y},A$} 
	\State Generate a random initial configuration $\V{c}$ with $\sum_{i}c_i=N\rho$
	\For {$a=1:N_{\mu}$} \Comment{Changing temperature}
   		\For {$t=1:\tau_{a} $} 	\Comment{Sampling at $\mu=\mu_a$}
	   		\For {$i=1:N$} 		\Comment{Extensive number of MC updates}
   				\State $\V{c}\lA {\rm MC_{PF}}(\V{c},\mu_a,\V{y},A)$  
	   		\EndFor
			\State \# Calculate energy $\epsilon_t=\epsilon(\V{c}_t|\V{y},A)$  
			of the current configuration $\V{c}_t=\V{c}$
  		\EndFor 
		\State \# Calculate the average energy $\epsilon_a=(1/\tau_a) \sum_{t=1}^{\tau_a} \epsilon_t$		
	\EndFor
	\State \Return $\V{c}$ 
\EndProcedure
\end{algorithmic}
\end{algorithm}
%%%%%%%%%%%%%%%%%%%%%%%%%%%%%%%
The lines marked with \# are not necessarily needed for SA, but have been inserted for later convenience. For \Rcode{SA}, we have a set of inverse temperature points $\{ \mu_a \}_{a}^{ N_{\mu} }$  arranged in ascending order $(0=)\mu_{1}<\mu_{2}<\cdots<\mu_{N_{\mu}}(\gg1)$ and the waiting times $\{\tau_a \}_{a}$ at those points. Hence, as the algorithm proceeds, the temperature of the system $T=1/\mu$ decreases step by step. It is known that, if the rate of decrease of the temperature obeys  
\be
T(t) > \frac{A(N)}{\log(t+2)}
\Leq{Geman}
\ee
for some time-independent constant $A(N)$, then the output of SA is guaranteed to be optimal~\cite{Geman:84}. This is a very slow schedule of decrease for $T$, and is generally overcautious so as to include the worst-case scenario. Faster schedules are known to work in practical situations. We report the results for such a rapid annealing below. 

%%%%%%%%%%%%%%%%%%%%%%%%%%%%%%%%%%%%%%%%%%%%%%%%%%%%%%
%%%%%%%%%%%%%%%%%%%%%%%%%%%%%%%%%%%%%%%%%%%%%%%%%%%%%%
%%%%%%%%%%%%%%%%%%%%%%%%%%%%%%%%%%%%%%%%%%%%%%%%%%%%%%
\section{Results}
We now present the results of SA according to \Req{P(c)}. System sizes of $N=100,200$, and $400$ will be examined. The annealing schedule is fixed as
\be    
\mu_a=\mu_0+r^{a-1}-1,~\tau_a=\tau,~(a=1,\cdots,100).
\Leq{schedule}
\ee
We set $\tau=5$, $\mu_0=10^{-8}$, and $r=1.1$ as default parameter values. Thus, the maximum value of $\mu$ is  $\mu_{100}\approx 1.3 \times 10^{4}$. 

To determine whether the SA process is proceeding well, we consider the lines marked with \# in \Rcode{SA}. If our schedule is sufficiently slow, the $\V{c}_t$ obtained during annealing are typical samples from \Req{P(c)}, and the values of physical quantities of typical samples should be very close (in the limit $N\to \infty$ almost surely identical) to the thermal averages. This implies the following relation 
\be
\epsilon_a=\frac{1}{\tau_a}\sum_{t}^{\tau_a}\epsilon(\V{c}_t) \approx \Ave{\epsilon(\V{c}|\V{y},A) }_{\mu_a},
\ee
where $\Ave{\cdots}_{\mu}$ denotes the average over \Req{P(c)} with the inverse temperature $\mu$. Fortunately, the right-hand side is analytically assessed in the present case with a random matrix dictionary $A$ in the $N\to \infty$ limit~\cite{Nakanishi:15,Obuchi:15}. Hence, we compare $\mc{E}_a$ with the analytically evaluated $\Ave{\mc{E}}_{\mu_a}$ to determine how well our annealing process follows the correctly distributed samples from \NReq{P(c)}. For clarity of comparison,  we take an average over a different $N_{\rm samp}=100$ samples of $\hat{\V{x}},\V{\xi},A$ in the numerical experiments. The error bar is given by the standard deviation among those samples divided by $\sqrt{N_{\rm samp}-1}$. 

As well as the distortion, we calculate the mean squared error (MSE) between the planted and inferred representations 
\be
\mc{M}(\V{c})=\frac{1}{N}||\hat{\V{x}}-\V{c}\circ  \V{x}(\V{c}) ||_2^2.
\ee
This metric provides direct information about the reconstruction of the planted solution. 

%%%%%%%%%%%%%%%%%%%%%%%%%%%%%%%%%%%%%%%%%%%%%%%%%%%%%%
%%%%%%%%%%%%%%%%%%%%%%%%%%%%%%%%%%%%%%%%%%%%%%%%%%%%%%
\subsection{Noiseless case}
Let us start with the noiseless case $\sigma_{\xi}^2=0$. In this subsection, we fix $\sigma_x^2=1$. 

The easy reconstruction region, where $\alpha$ is sufficiently larger than $\hat{\rho}$, is a good starting point. \Rfig{eps_M-easy} plots  $\epsilon$ (left, middle) and the MSE $\mc{M}$ (right) against the temperature for $\alpha=0.8,\rho=0.4$, and $\hat{\rho}=0.2$.
%%%%%%%%%%%%%%%%%%%%%
\begin{figure}[htbp]
\begin{center}
\includegraphics[width=0.325\columnwidth,height=0.28\columnwidth]{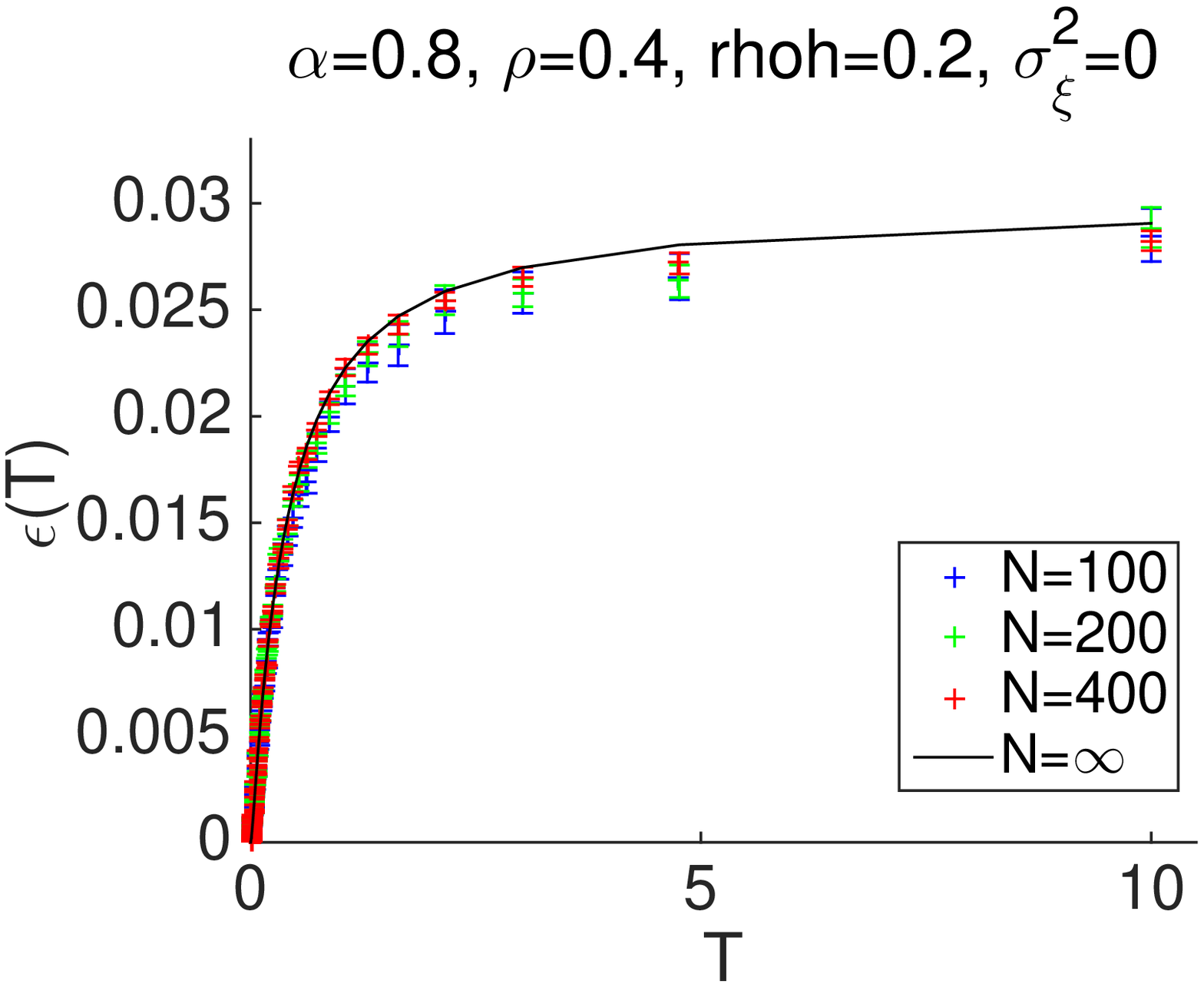}
\includegraphics[width=0.325\columnwidth,height=0.28\columnwidth]{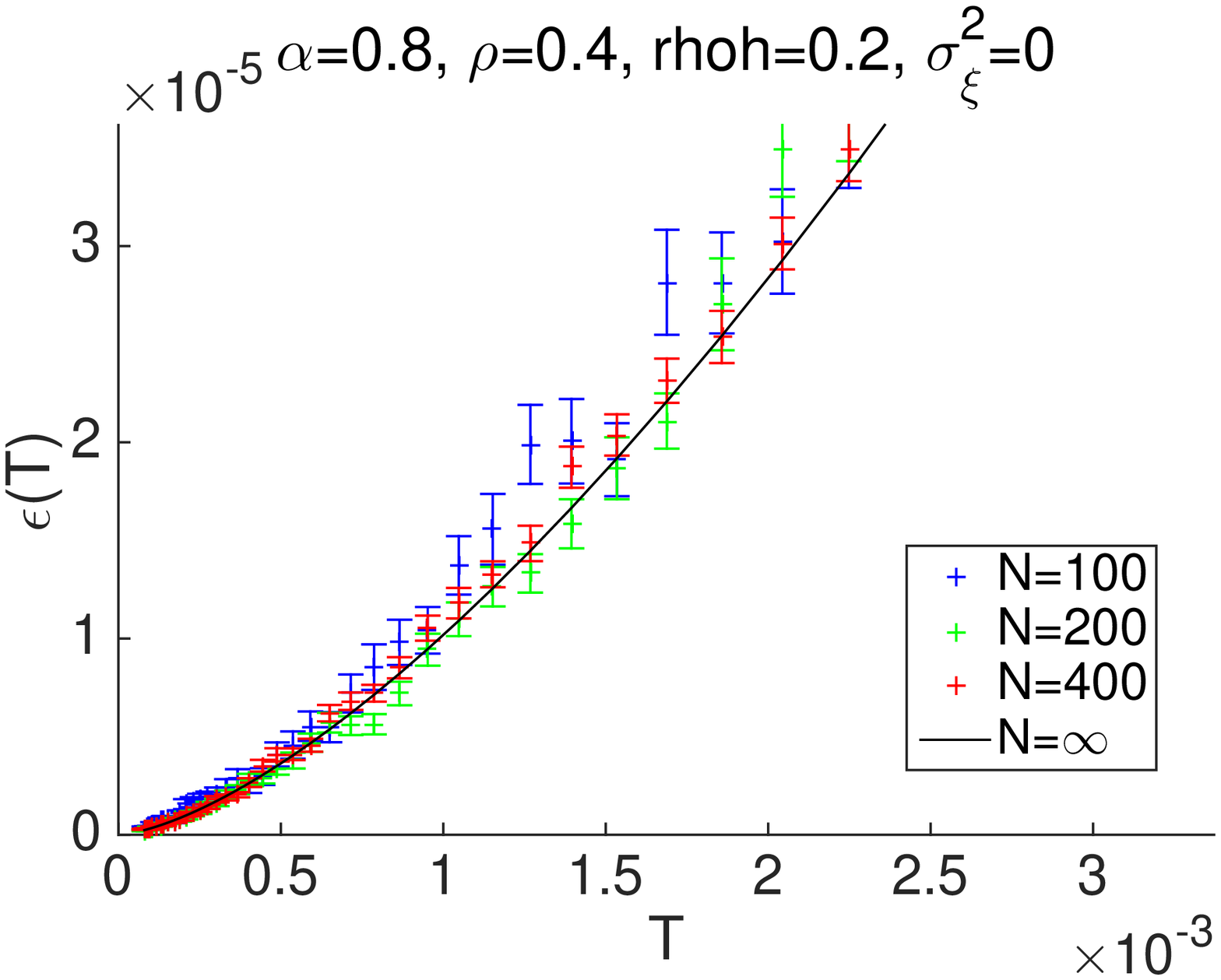}
\includegraphics[width=0.325\columnwidth,height=0.28\columnwidth]{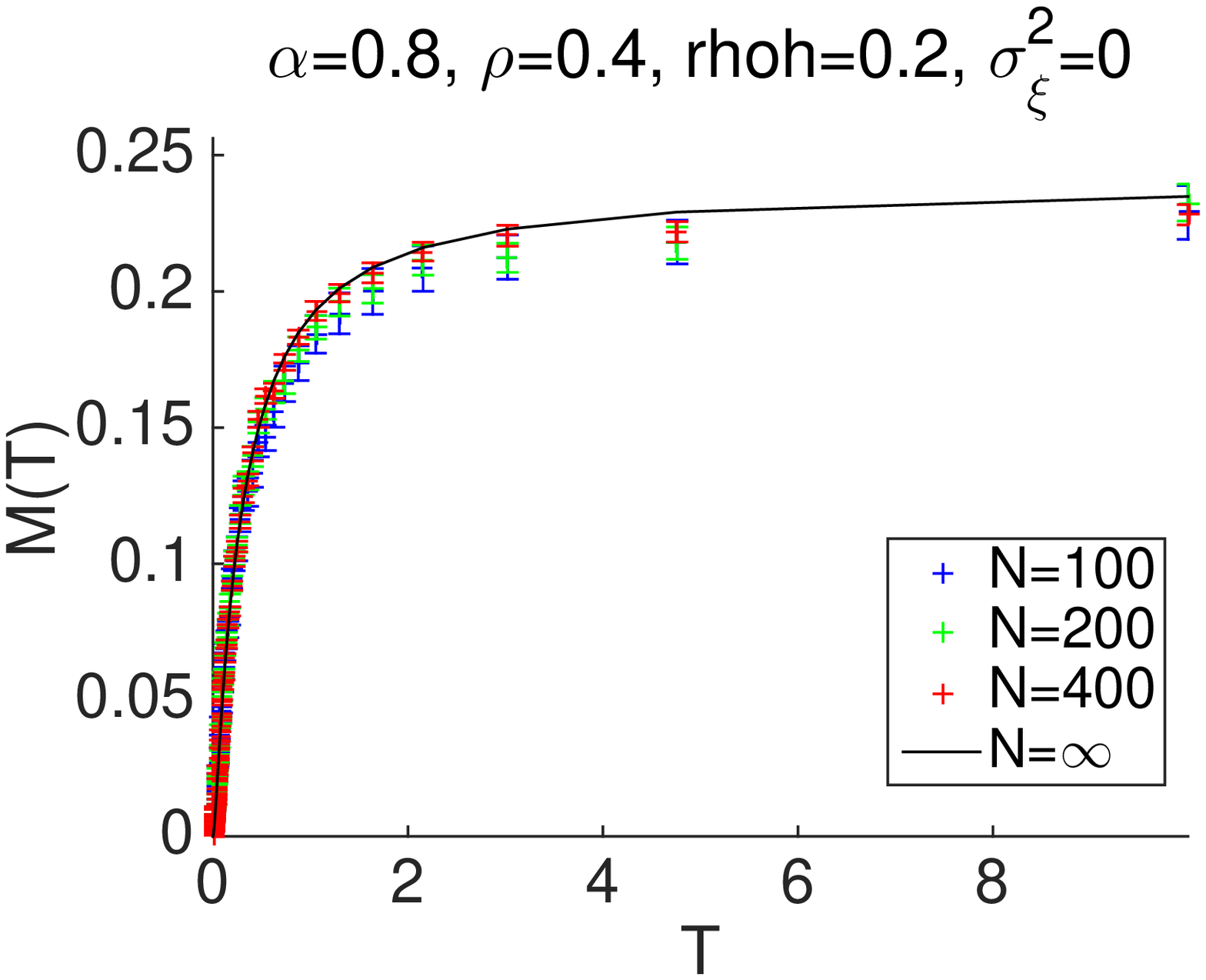}
\end{center}
\caption{Distortion (left, middle) and MSE (right) plotted against temperature $T=1/\mu$ for $\alpha=0.8,\rho=0.4$, and $\hat{\rho}=0.2$. The middle panel is a magnified view of the left panel for a small region of $T$. The solid black curve shows the analytical values, and the color plots give the numerical results. Our numerical results clearly reproduce the correct average values over  \Req{P(c)}.}
\Lfig{eps_M-easy}
\end{figure}
%%%%%%%%%%%%%%%%%%%%%
The numerical results show a fairly good agreement with the analytical curve. This means our SA algorithm  follows the equilibrium state up to the zero-temperature limit reasonably well, even though the annealing defined by \Req{schedule} is very rapid. The MSE goes to zero as $T$ decreases, and so the planted solution is correctly reproduced. 

Next, we consider harder cases. It is known that the properties of the problem drastically change as $\alpha$ gets closer to $\hat{\rho}$. The phase diagram in \Rfig{PD} demonstrates this. 
%%%%%%%%%%%%%%%%%%%%%
\begin{figure}[ht]
\begin{center}
\includegraphics[width=0.4\columnwidth]{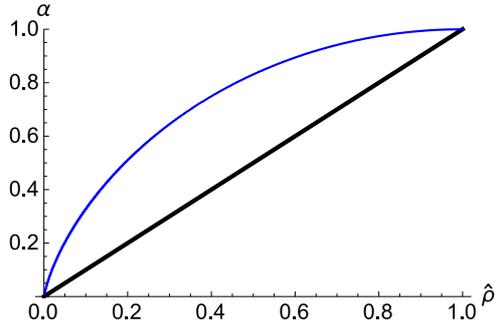}
\end{center}
\caption{Phase diagram describing typical reconstruction limits in the noiseless case~\cite{Kabashima:09}. The straight line $\hat{\rho}=\alpha$ is the limit attained with \Req{optimization}, whereas the curve is the limit for the relaxed problem in which the $\ell_0$ norm in \Req{optimization} is replaced with the $\ell_1$ norm. Above these boundaries, the solutions of corresponding optimization problems reconstruct the planted solution $\hat{\V{x}}$. }
\Lfig{PD}
\end{figure}
%%%%%%%%%%%%%%%%%%%%%
The curve is the reconstruction limit of the $\ell_1$-relaxed version of \NReq{optimization}, which is employed in many realistic cases and has considerable importance. Hence, we first examine the behavior of SA below this $\ell_1$ reconstruction limit. \Rfig{eps_M-hard1} plots  $\epsilon$ and $\mc{M}$ against $T$ for $\alpha=0.8,\rho=0.55$, and $\hat{\rho}=0.5$, where we are below the $\ell_1$ reconstruction limit (the boundary is located at $\alpha_c(\hat{\rho}=0.5)\approx 0.831$).   
%%%%%%%%%%%%%%%%%%%%%
\begin{figure}[htbp]
\begin{center}
\includegraphics[width=0.48\columnwidth]{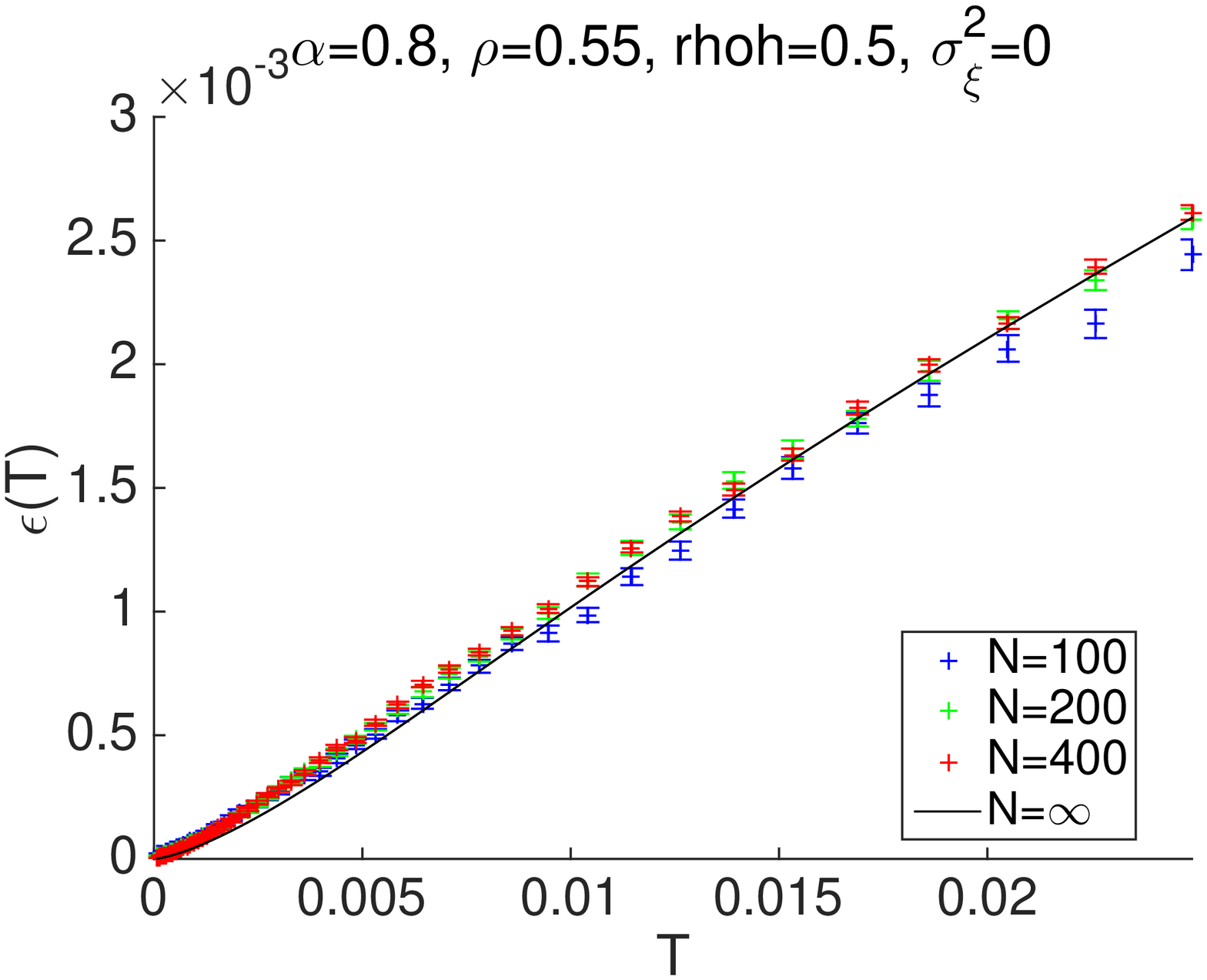}\includegraphics[width=0.48\columnwidth]{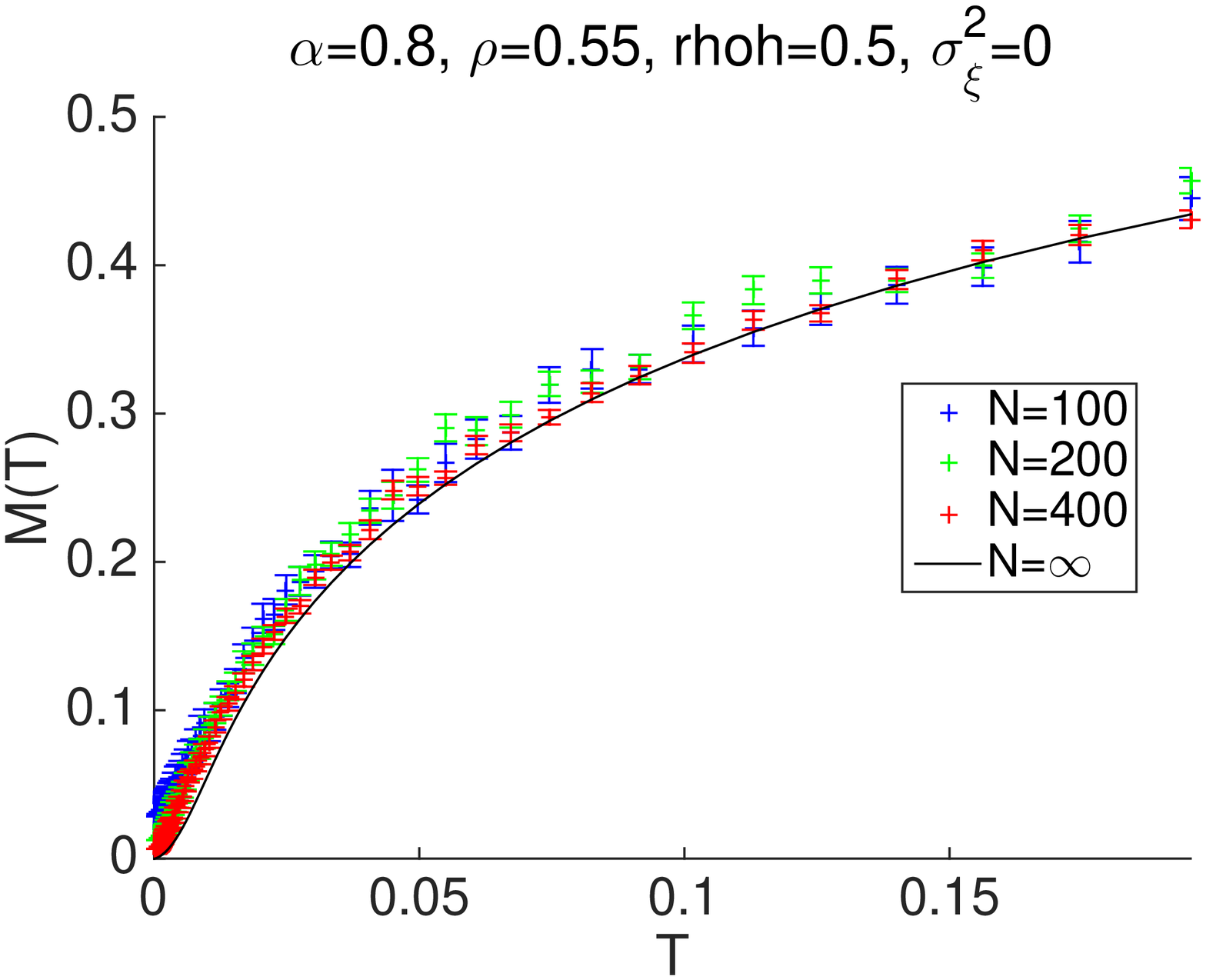}
\end{center}
\caption{Distortion (left) and MSE (right) plotted against temperature for $\alpha=0.8,\rho=0.55$, and $\hat{\rho}=0.5$ (below the $\ell_1$ boundary). Our numerical results accord with the analytical result (black line) and achieve a perfect reconstruction of $\hat{\V{x}}$. }
\Lfig{eps_M-hard1}
\end{figure}
%%%%%%%%%%%%%%%%%%%%%
The MSE vanishes as $T$ decreases, meaning that $\hat{\V{x}}$ is perfectly reconstructed. Hence, SA can outperform the $\ell_1$ method of reconstruction, even under the present rapid schedule. 

In harder situations, SA cannot always give a perfect reconstruction, even though it exists. \Rfig{eps_M-hard2} shows plots of $T$-$\epsilon$ and $T$-$\mc{M}$ for $\alpha=0.75,\rho=0.65$, and $\hat{\rho}=0.5$. 
%%%%%%%%%%%%%%%%%%%%%
\begin{figure}[htbp]
\begin{center}
\includegraphics[width=0.48\columnwidth]{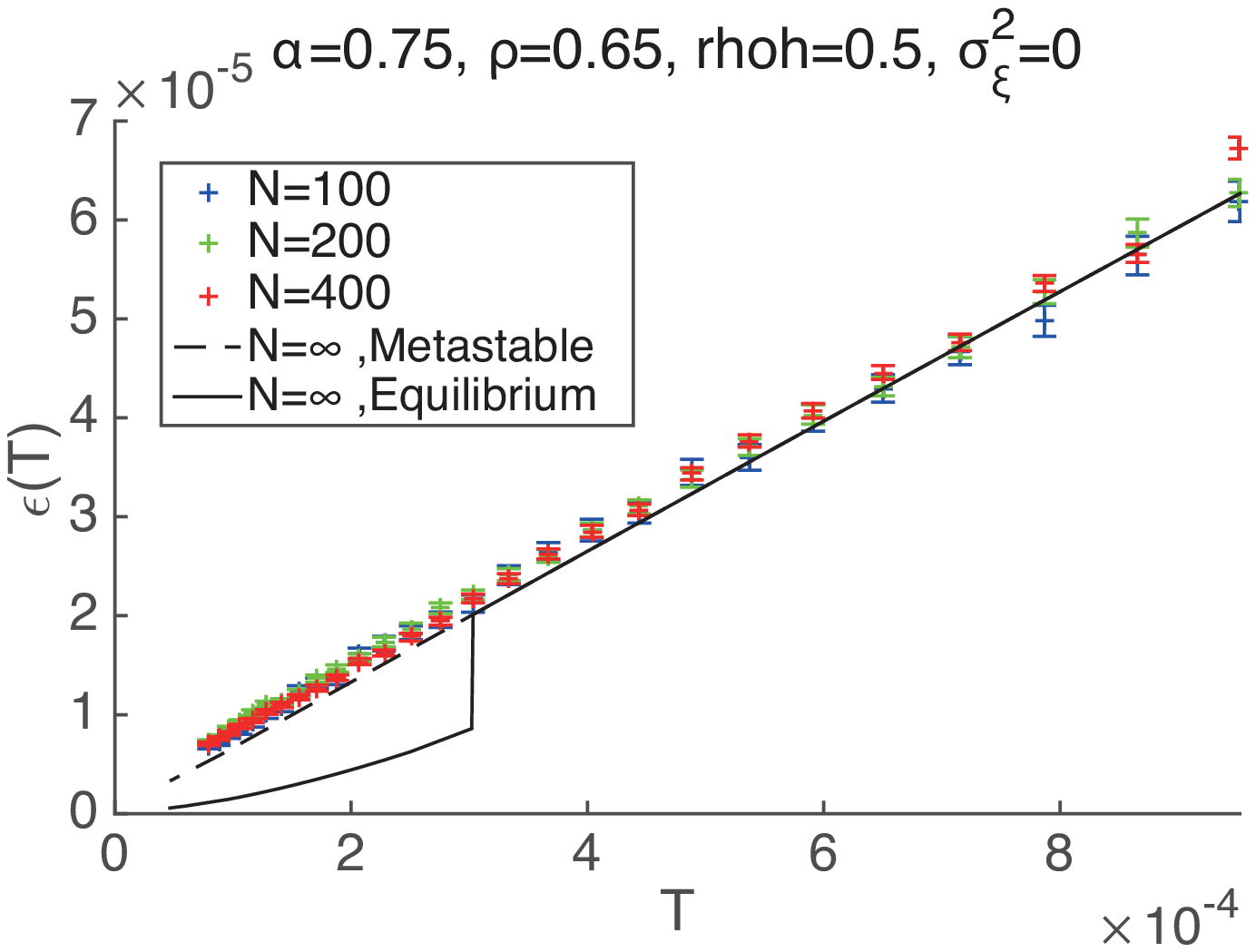}
\includegraphics[width=0.48\columnwidth]{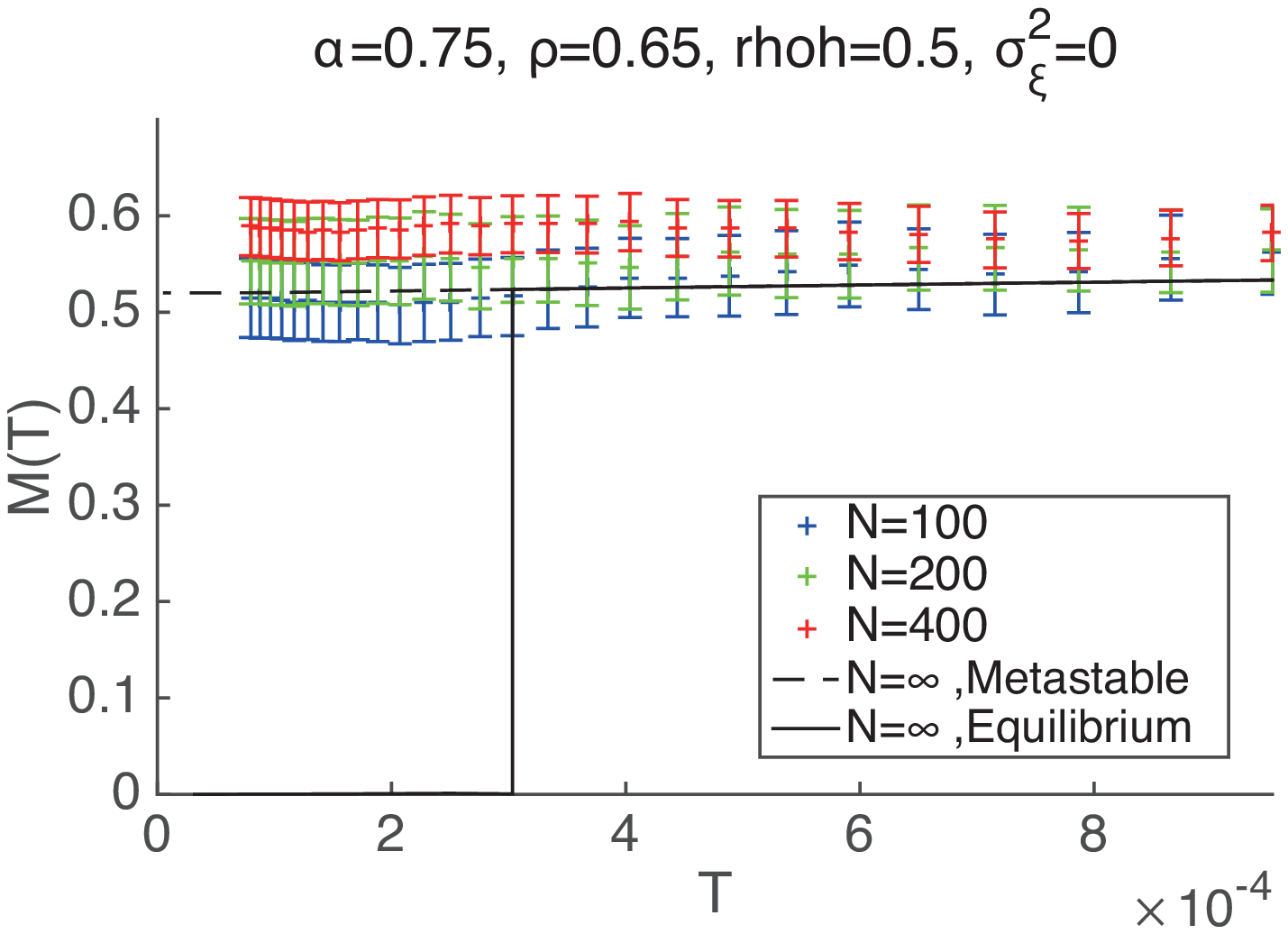}
\end{center}
\caption{Distortion $\epsilon$ (left) and MSE $\mc{M}$ (right) plotted against temperature for $\alpha=0.75,\rho=0.65$, and $\hat{\rho}=0.5$. The analytical result shows a first-order phase transition around $T_c \approx 3\times 10^{-4}$, and the equilibrium state (black solid line) suddenly decreases for both $\epsilon$ and $\mc{M}$. However, a metastable state, which continues analytically to the equilibrium state for $T>T_c$, survives below $T_c$, and the numerical results follow this; hence, the planted solution is not reached. }
\Lfig{eps_M-hard2}
\end{figure}
%%%%%%%%%%%%%%%%%%%%%
In this case, there are two stable thermodynamic states at low temperatures: one is connected to the planted solution and is the true equilibrium state in the zero temperature limit, whereas the other is metastable at  low temperatures, but is the dominant equilibrium state at higher temperatures~\cite{Obuchi:15}. There is a first-order phase transition at a critical temperature $T_c\approx 3\times 10^{-4}$. Through the SA process, the system state follows the equilibrium up to $T>T_c$. After the transition, the equilibrium state changes drastically,  but the system cannot follow such a jump. Instead, the system remains in the same metastable state  in $T<T_c$. Hence, SA cannot find the planted solution in this case, as clearly seen in the non-vanishing MSE $\mc{M}$ of \Rfig{eps_M-hard2}. Of course, if $\tau$ is large enough, SA can eventually find the planted solution as proved in \cite{Geman:84}. In the presence of the first-order phase transition, however, the required $\tau$ to do this is scaled with the system size $N$ and rapidly grows as $N$ increases, which prevents reaching the optimal solution in practical times. 

Although the system is in the metastable state, the distortion $\epsilon$ seems to go to zero in \Rfig{eps_M-hard2}, but unfortunately this is not the case. If we go lower temperatures, the value of $\epsilon$ will get stuck at a certain critical temperature and remains a constant below it. This is a freezing transition of this metastable state: the entropy of it becomes zero at the transition temperature. With the present parameters, our analytical solution provides the transition temperature as $T\approx 1.3 \times 10^{-7}$ and the constant value of distortion is $\epsilon=8.6 \times 10^{-9}$ which is the achievable limit by our SA algorithm in practical times. This limiting value is not exactly zero although it is still negligibly small. For reference, we have plotted the ``free energy'' $g=(1/N)\log G$ and the entropy in \Rfig{g-hard2} for the present parameter values. 
%%%%%%%%%%%%%%%%%%%%%
\begin{figure}[htbp]
\begin{center}
\includegraphics[width=0.48\columnwidth]{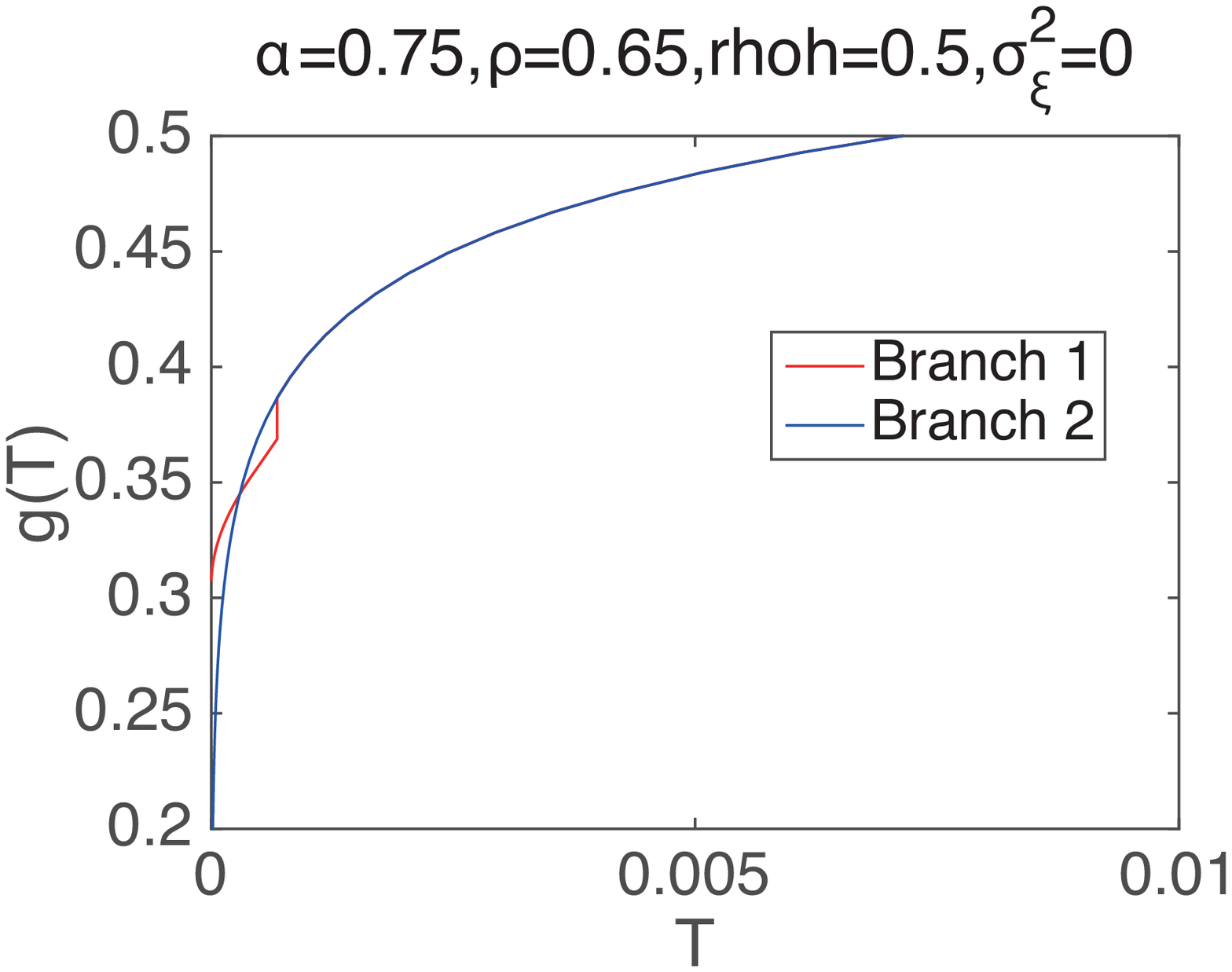}
\includegraphics[width=0.48\columnwidth]{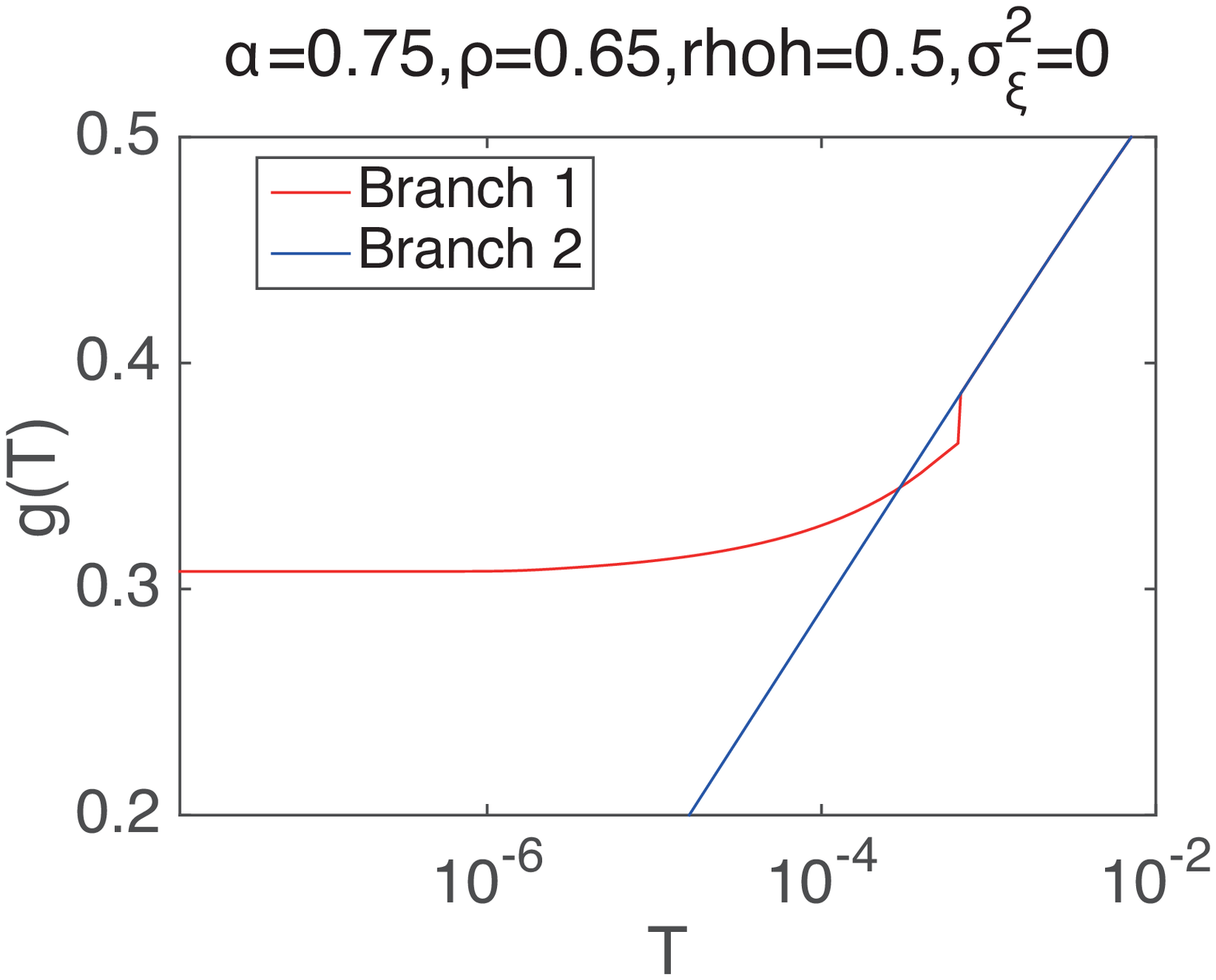}
\includegraphics[width=0.48\columnwidth]{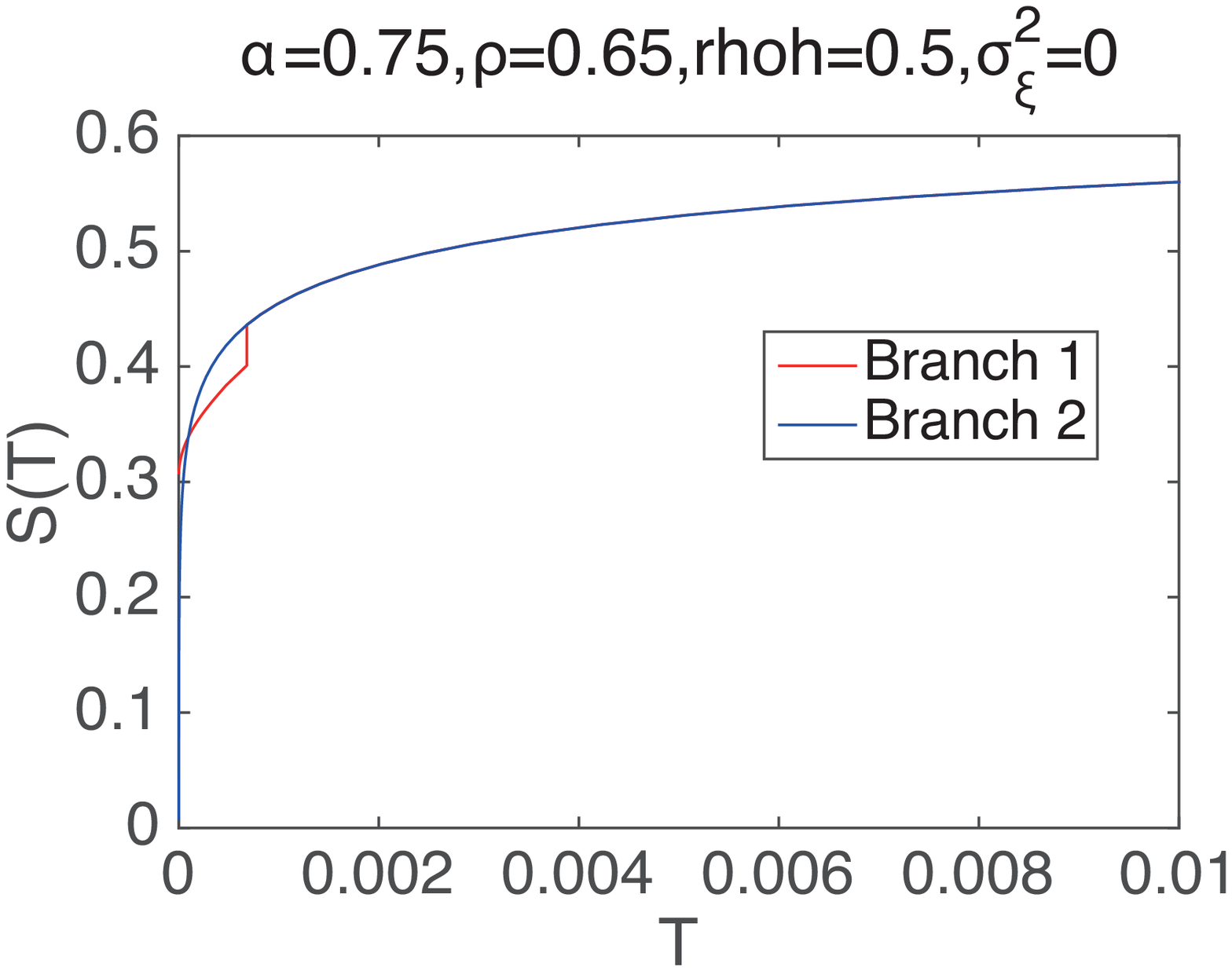}
\includegraphics[width=0.48\columnwidth]{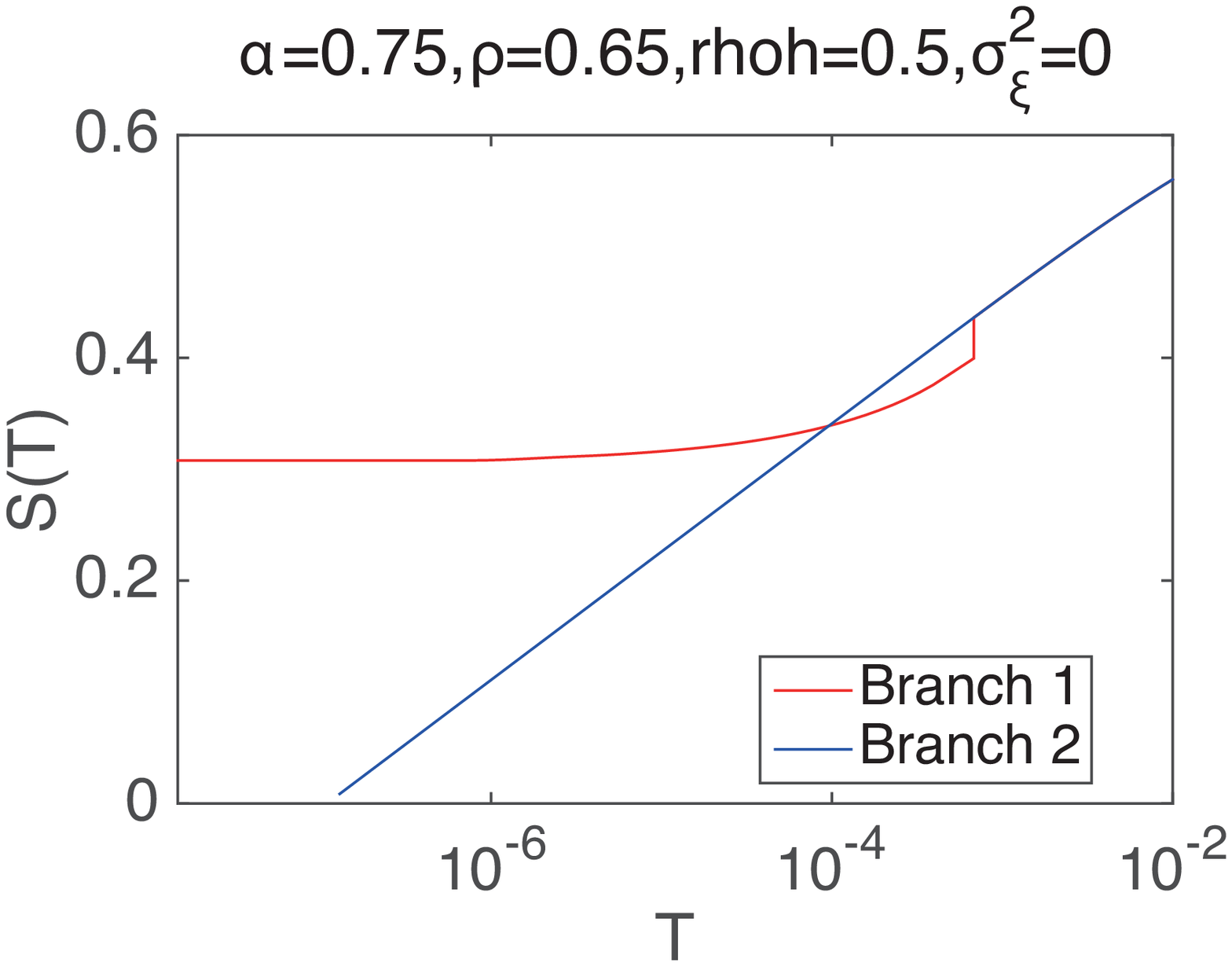}
\end{center}
\caption{ Free energy $g=(1/N)\log G$ (upper) and  entropy (lower) plotted against temperature for $\alpha=0.75,\rho=0.65$, and $\hat{\rho}=0.5$. Right panels are the same plots as the left panels but with a semi-logarithmic scale. Two different branches are shown, and the same colors denote the same branches in all panels. A phase transition (where two branches of $g$ cross) can be observed at $T_{c}\approx 3\times 10^{-3}$. The larger branch yields the equilibrium state. The entropy of the Branch 2 vanishes at a finite value of $T$, as clearly shown in the lower-right panel. Branch $1$, the red solid curve, is connected to the planted solution.
 }
\Lfig{g-hard2}
\end{figure}
%%%%%%%%%%%%%%%%%%%%%

Finally, we examined the effect of the parameter $\rho$, which can be set to any value for a given $\V{y}$ and $A$. \Rfig{eps_M-hard3} is the counterpart of \Rfig{eps_M-hard2} with common values of $(\alpha,\hat{\rho})=(0.75,0.5)$ and a different value of $\rho=0.55$. 
%%%%%%%%%%%%%%%%%%%%%
\begin{figure}[htbp]
\begin{center}
\includegraphics[width=0.48\columnwidth]{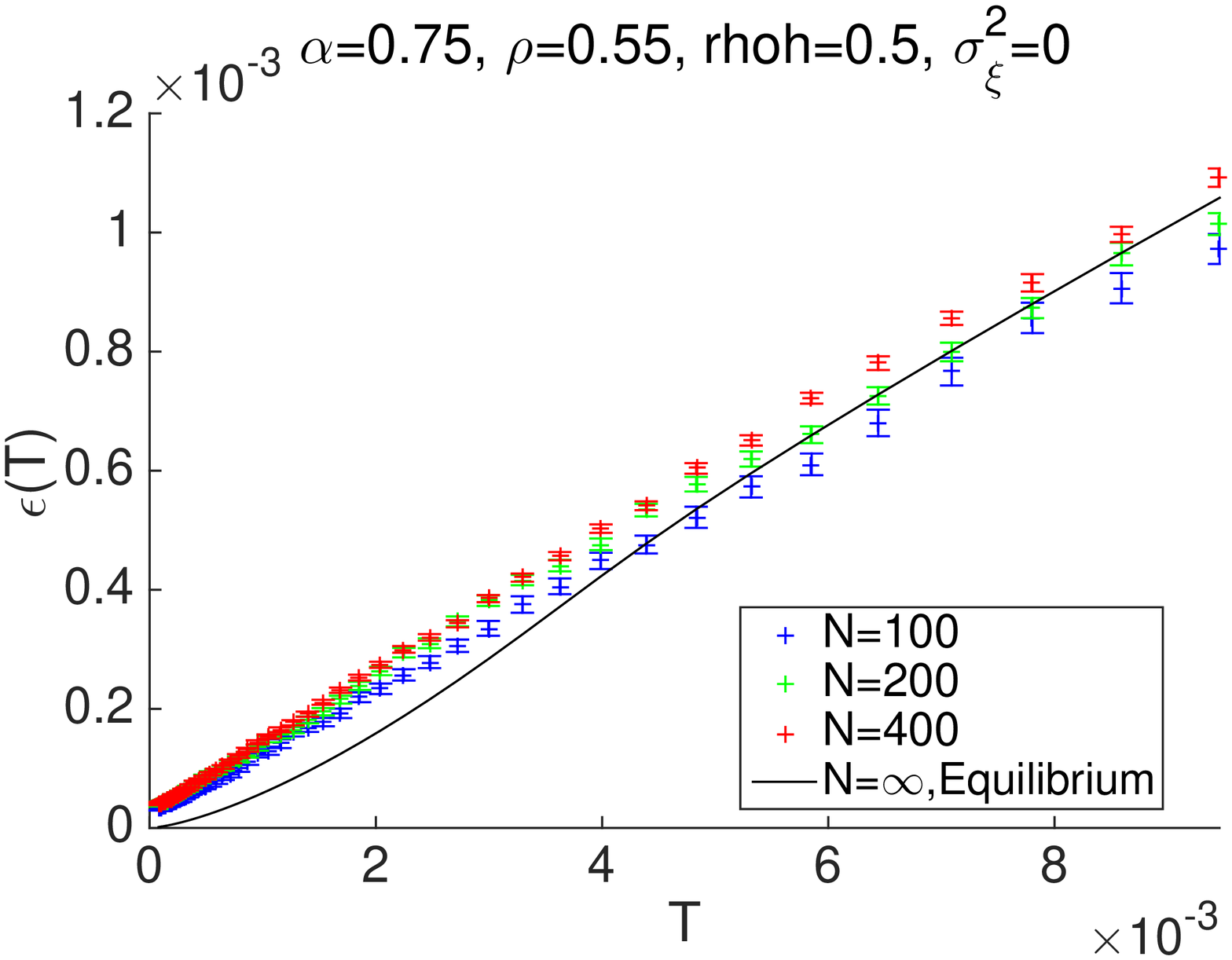}
\includegraphics[width=0.48\columnwidth]{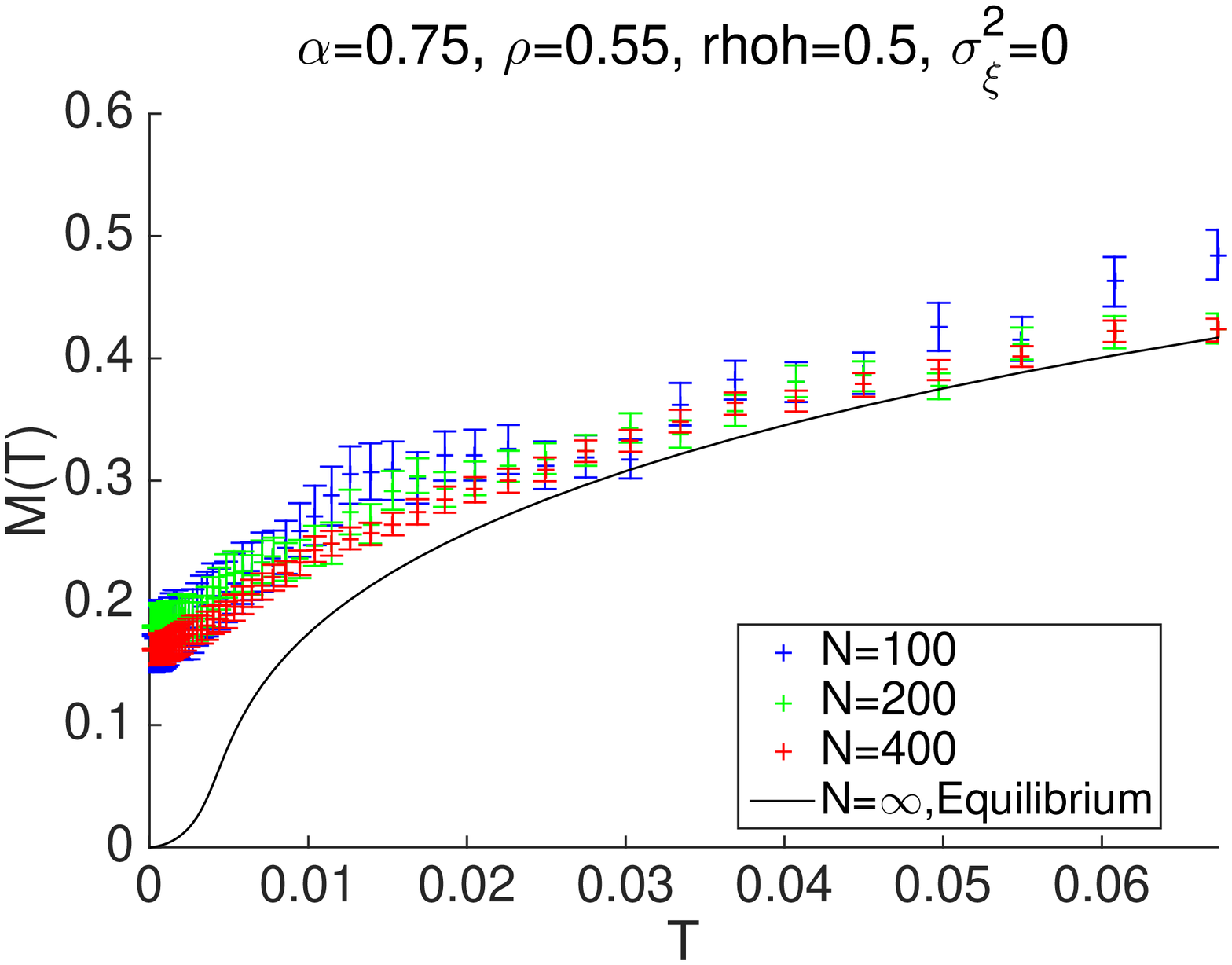}
\includegraphics[width=0.48\columnwidth]{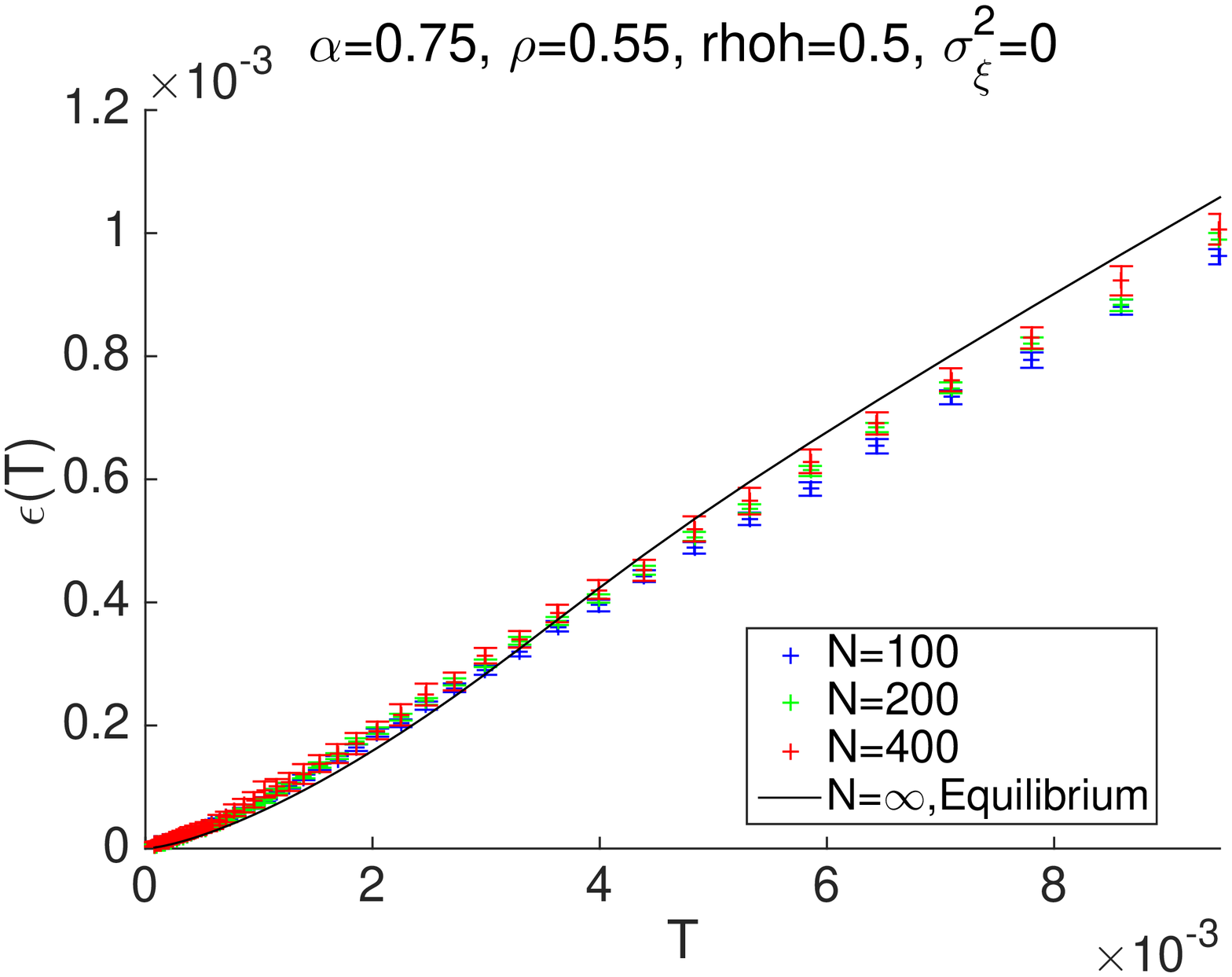}
\includegraphics[width=0.48\columnwidth]{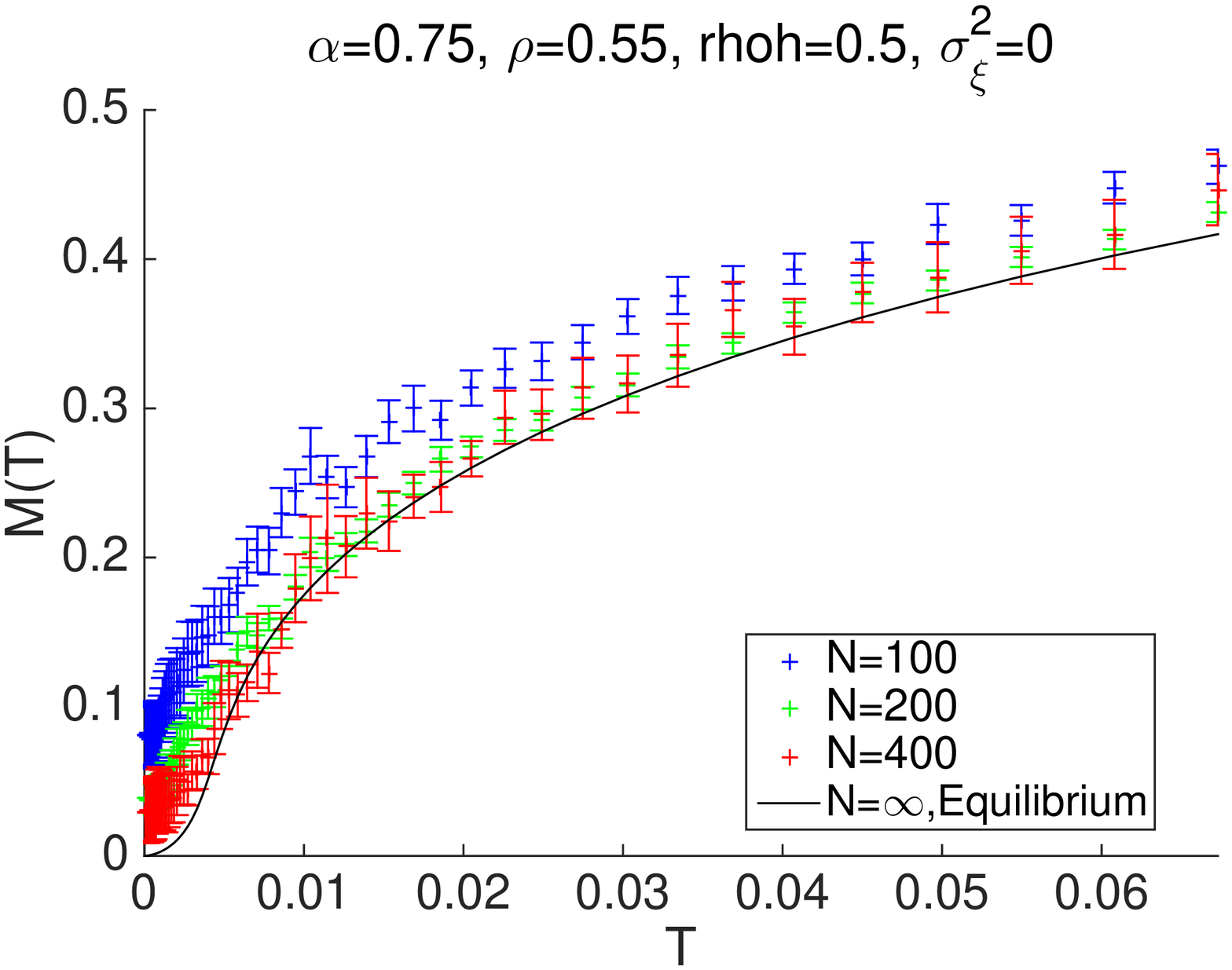}
\end{center}
\caption{Distortion $\epsilon$ (left) and MSE $\mc{M}$ (right) plotted against temperature for $\alpha=0.75,\rho=0.55$, and $\hat{\rho}=0.5$. In contrast to \Rfig{eps_M-hard2} with $\rho=0.65$, there is no phase transition. The upper panels correspond to a rapid schedule with $\tau=5$, whereas the lower ones correspond to $\tau=100$. At low temperatures, the rapid case does not follow the equilibrium state, whereas the slower one is well matched.    }
\Lfig{eps_M-hard3}
\end{figure}
%%%%%%%%%%%%%%%%%%%%%
In contrast to \Rfig{eps_M-hard2}, the metastable state is absent. We commonly observed the suppression of this metastable state as $\rho$ decreased~\cite{Obuchi:15}. Thus, SA was naively expected to yield the planted solution in the zero temperature limit. The upper panels of \Rfig{eps_M-hard3} represent the rapid schedule $\tau=5$, and exhibit a clear deviation from the equilibrium state at low temperatures. A slower schedule, corresponding to $\tau=100$, is represented in the lower panels, and these numerical data are consistent with the analytical curve of the equilibrium state. The present choice of parameters is in the harder region, so it is unsurprising that the rapid schedule $\tau=5$ is not sufficiently slow, although it was for \Rfigss{eps_M-easy}{eps_M-hard1}. Note that this increase in $\tau$ is qualitatively different from that caused by the emergence of the metastable state in \Rfig{eps_M-hard2}. The latter is macroscopic, namely, the required waiting time $\tau$ increases significantly as $N$ grows to reach the planted solution. 

In summary, SA outperforms the $\ell_1$ result and can correctly find the planted solution in a wider region of parameter space, even under our rather rapid rate of decrease of $T$. However, for harder regions where $\alpha$ is close to $\hat{\rho}$, the phase space is separated into two distinctive states. The state that is not connected to the planted solution becomes the equilibrium state in the high-temperature region. This prevents SA from correctly finding the planted solution, as the system becomes trapped in the wrong state. Tuning $\rho$ may be a crucial factor in overcoming this. As long as $\rho>\hat{\rho}$, smaller values of $\rho$ are better. This is because the emergence of the metastable state tends to be suppressed as $\rho$ becomes smaller, but such fine tuning requires {\it a priori} knowledge about the value of $\hat{\rho}$, which is not available in most situations. This problem is of considerable importance, and requires further consideration. 

If we do not insist on finding the planted solution, even the metastable state may be desirable, as it shows small distortion. In this case, larger values of $\rho$ are better, because they yield smaller values of distortion, though the compression ratio decreases as $\rho$ increases. Hence, the value of $\rho$ should be chosen according to the requirements of the information processing application. 

%%%%%%%%%%%%%%%%%%%%%%%%%%%%%%%%%%%%%%%%%%%%%%%%%%%%%%
%%%%%%%%%%%%%%%%%%%%%%%%%%%%%%%%%%%%%%%%%%%%%%%%%%%%%%
\subsection{Case without planted solutions}
Next, we examine the opposite case with $\sigma_{\xi}^2=1$ and $\sigma_x^2=0$. There is now no planted solution,  and we focus solely on how small $\epsilon$ becomes. For comparison, we present a number of values of $\epsilon$ achieved by different methods. The symbol ell$_1$ denotes the $\ell_1$ method in~\cite{Nakanishi:15}, in which the $\ell_1$-relaxed version of \Req{optimization} is solved and the resultant solution of $\V{x}$ is inserted in \Req{epsilon(x)}. Similarly, the symbol ell$_1$+LS corresponds to the method in~\cite{Nakanishi:15}, which gives $\epsilon$ obtained by \Req{epsilon(c)}, but with the substituted support $\V{c}$ determined by solving the $\ell_1$-relaxed version of \Req{optimization}. The symbol OMP denotes the results given by orthogonal matching pursuit~\cite{Pati:93,Davis:94}. The OMP result is not obtained by analytical methods, but by numerical experiments with the same parameters as SA for $N=400$.

\Rfig{eps-T1} plots $\epsilon$ against temperature $T$. Note that the thermal average of the distortion stops decreasing at a certain value of $T$.  Below this, $\epsilon$ remains constant, giving the achievable limit of the value of distortion in the present case~\cite{Nakanishi:15}, as seen in the analytical result (black solid curve).
%%%%%%%%%%%%%%%%%%%%%
\begin{figure}[ht]
\begin{center}
\includegraphics[width=0.48\columnwidth]{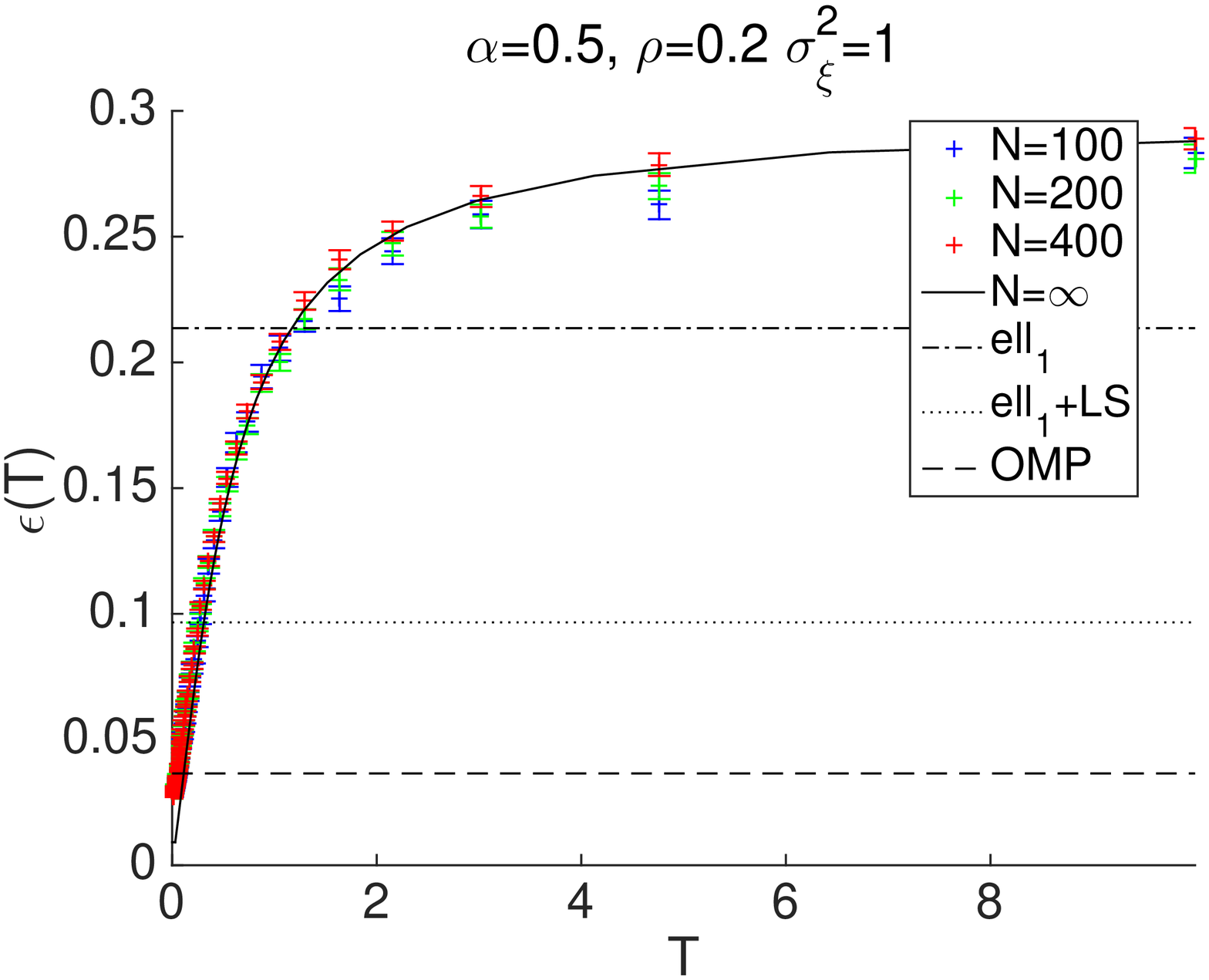}
\includegraphics[width=0.48\columnwidth]{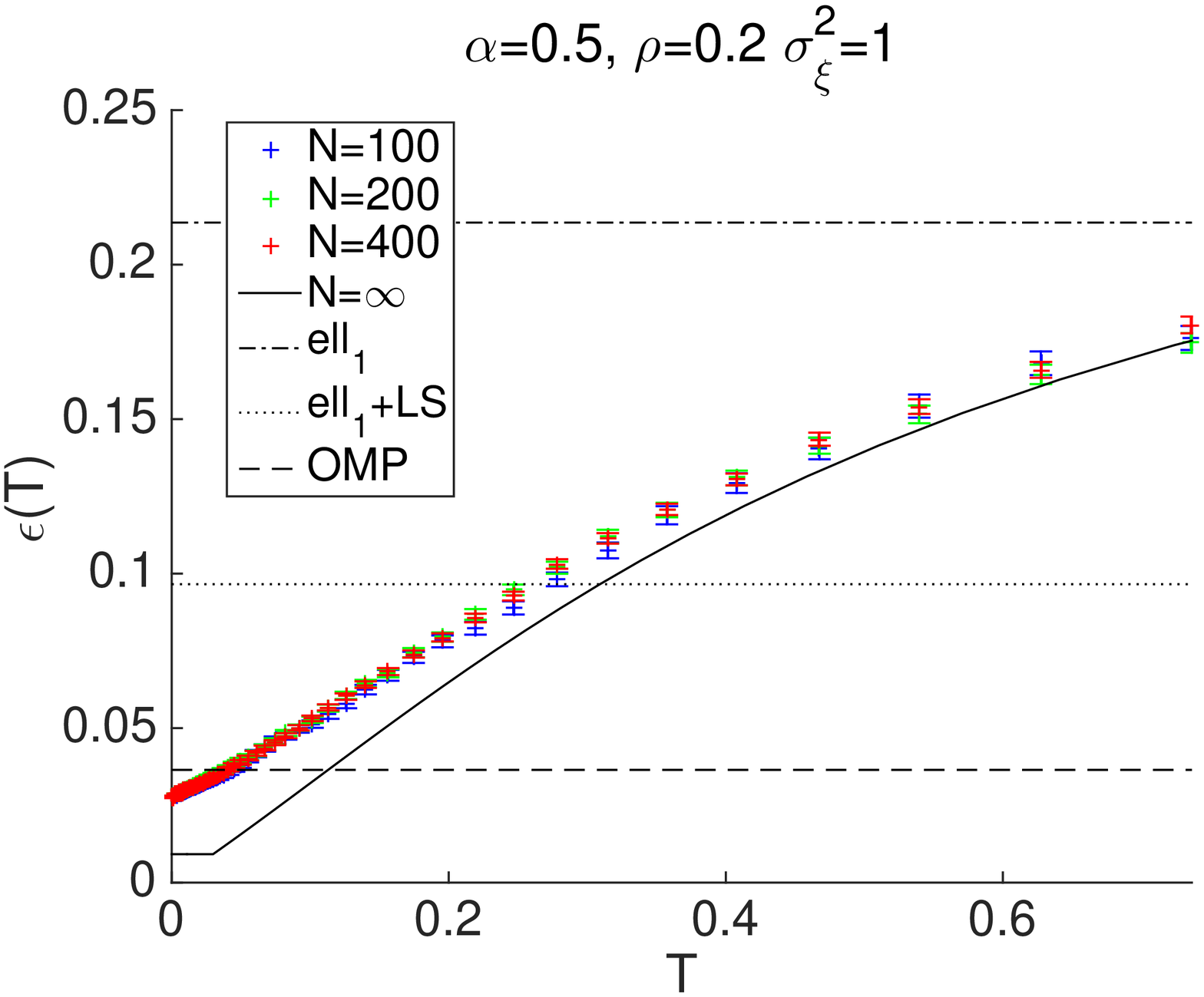}
\end{center}
\caption{Distortion plotted against temperature for $\alpha=0.5,\rho=0.2$ for the case without a planted solution. The right panel is a magnified view of the left one at low temperatures. The black solid curve represents the analytical value of $\epsilon$. Values of $\epsilon$ given by different algorithms are also displayed as horizontal lines with different symbols. See the main text for details.
}
\Lfig{eps-T1}
\end{figure}
%%%%%%%%%%%%%%%%%%%%%
Over a longer temperature range (left panel), the numerical results seem to agree well with the analytical curve, but over a tighter range of low temperatures (right panel), the numerical data exhibit a systematic deviation from the analytical curve. This is likely to be a result of the rapid nature of our annealing schedule. A slower schedule will improve the achievable value of distortion, as in \Rfig{eps_M-hard3}. Regardless, we can see that the SA result is already better than the values achieved by the other methods, demonstrating the effectiveness of SA. For reference, the distortion values obtained by $\ell_1$, $\ell_1+{\rm LS}$, OMP, and SA at $\mu=\mu_{100}\approx 1.3\times 10^{4}$ and $N=400$ were $\epsilon_{\ell_1}=0.214,\epsilon_{\ell_1+{\rm LS}}= 0.0966,\epsilon_{{\rm OMP}}=0.0365\pm  6.7 \times 10^{-4}$, $\epsilon_{\rm SA}=0.0272 \pm  6.2\times 10^{-4}$, respectively. The achievable limit was $\epsilon_{0}=0.00919$.

We also examined other parameter values, but the qualitative behavior was the same as that of \Rfig{eps-T1}, and so the results are not shown here. 

%%%%%%%%%%%%%%%%%%%%%%%%%%%%%%%%%%%%%%%%%%%%%%%%%%%%%%
%%%%%%%%%%%%%%%%%%%%%%%%%%%%%%%%%%%%%%%%%%%%%%%%%%%%%%
%%%%%%%%%%%%%%%%%%%%%%%%%%%%%%%%%%%%%%%%%%%%%%%%%%%%%%
\section{Discussion}

%%%%%%%%%%%%%%%%%%%%%%%%%%%%%%%%%%%%%%%%%%%%%%%%%%%%%%
%%%%%%%%%%%%%%%%%%%%%%%%%%%%%%%%%%%%%%%%%%%%%%%%%%%%%%
\subsection{Computation time and its order}
For reference, we give the actual computation times for one run of SA under the schedule in \Req{schedule}: for $\alpha=0.5$ and $\rho=0.2$, approximately $6, 15$, and $38$ seconds were required for $N=100,200$, and $400$, respectively. This experiment was performed on a 1.7 GHz Intel Core i7 with two CPUs using MATLAB\textsuperscript{\textregistered}. 

As well as these practical results, we can estimate the order of the computation time. Formally, this can be written as $O(N_{\mu}\tau N N_{\rm MC})$, the last factor of which is the computational cost of each MC update. The most expensive part is the matrix multiplication and inversion required to calculate the energy. If we use simple multiplication and Gauss elimination in the inversion process for each step, $N_{\rm MC}=O(M(N\rho)^2+(N\rho)^3)$. However, we have employed pair flipping in each update of the sparse weights, meaning that the change in the relevant matrices in each move is small and successive. This implies that we can reduce the computation time by successively updating those matrices while employing the matrix inversion formula. 

Provided that $G_{t+1}$ is decomposed as
\be
G_{t+1}=
\left(
\begin{array}{cc}
G_t  &  \V{g}_{t+1}     \\
 \V{g}_{t+1}^{\rm T} & g_{t+1}      
\end{array}
\right),
\ee
 the matrix inversion formula gives
\be
 G_{t+1}^{-1}=
\left(
\begin{array}{cc}
G_t^{-1}+\gamma_{t+1} G_t^{-1}\V{g}_{t+1}\V{g}_{t+1}^{\rm T}G_t^{-1}  &  -\gamma_{t+1} G_{t}^{-1} \V{g}_{t+1}     \\
 (-\gamma_{t+1} G_{t}^{-1} \V{g}_{t+1})^{\rm T} & \gamma_{t+1}      
\end{array}
\right)
\equiv
\left(
\begin{array}{cc}
U_{t} &  \V{u}_{t+1}     \\
  \V{u}^{\rm T}_{t+1} & u_{t+1}      
\end{array}
\right)
\equiv
U_{t+1},
\Leq{inversion}
\ee
where $\gamma_{t+1}=g_{t+1}-\V{g}_{t+1}^{\rm T}G_{t}^{-1} \V{g}_{t+1}$. We use this as follows. Write $G_{t+1}$ as $(\tilde{A}^{\rm T}(\V{c})\tilde{A}(\V{c}))$, and assume that we are given both $G_{t+1}$ and $G_{t+1}^{-1}=U_{t+1}$. First, we treat the deletion part of the MC move, $c_i=1 \to c'_{i}=0$. We want to calculate $G_{t}^{-1}$ from $G_{t+1}^{-1}$. This is given by 
\be
G_{t}^{-1}=U_{t}-\V{u}_{t+1}\V{u}^{\rm T}_{t+1}/u_{t+1},
\ee
which has a computational cost of $O((N\rho)^2)$. The corresponding $G_t$ is obtained by deleting the $i$th column and row from $G_{t+1}$. Next, we move to the addition part $c_j=0 \to c'_{j}=1$. We now have $G_{t}$ and $G_{t}^{-1}$. Extending the matrix, $G_{t}\to G_{t+1}$, involves adding an appropriate column and row to $G_{t}$. The $k$th component of the added column vector can be calculated as $\V{g}_{t+1}(k)=\sum_{l=1}^{M}A(l,k)A(l,j)$, where $k$ runs over the indices of ${\rm ONES}$, and thus the computational cost is $O(MN\rho)$. Similarly, we can calculate $g_{t+1}=\sum_{l=1}^{M}A(l,j)A(l,j)$, and hence the computational cost of calculating  $G_{t+1}$ is $O(MN\rho)$. Now, we can easily calculate $G_{t+1}^{-1}$ by \Req{inversion} from $G_{t}^{-1}$, $\V{g}_{t+1}$, and $g_{t+1}$. The computational cost of this operation is $O((N\rho)^2)$. This completes the successive update of $G=\tilde{A}^{\rm T}\tilde{A}$ and $G^{-1}$. 

In summary, the factor $N_{\rm MC}$ can be reduced to  
\be
N_{\rm MC}=O((N\rho)^2+MN\rho)=O(N^2\alpha \rho),~(M=N\alpha \geq N\rho).
\ee
Thus, the total computational cost of our SA algorithm is $O(\alpha \rho \tau N_{\mu}N^3)$. As long as $N_{\mu}$ does not scale with the size of the system, we have only third-order dependence on system-size, which is comparable to that of  versatile convex optimization solvers used in the $\ell_1$-relaxed version of the sparse approximation problem. Hence, our SA algorithm can solve the sparse approximation problem at a computational cost that is of the same order as the $\ell_1$-relaxed version, without any need to relax the problem. 

%%%%%%%%%%%%%%%%%%%%%%%%%%%%%%%%%%%%%%%%%%%%%%%%%%%%%%
%%%%%%%%%%%%%%%%%%%%%%%%%%%%%%%%%%%%%%%%%%%%%%%%%%%%%%
\subsection{Advantages, disadvantages, and possible extensions}
Our study indicates that SA reliably determines a solution with small distortion both in the presence and absence of noise, and has a reasonable computational cost. As long as noise and the metastable state are absent, the solution identified by SA is approximately equal to the planted solution. These findings encourage the use of SA in practical applications of the sparse approximation problem. 

To conclude, we summarize the advantages  and disadvantages of the present SA algorithm.
%%%%%%%%%%%%%%%%%%%%%%%%
\paragraph{Advantages:}
\begin{itemize}
\item{Easy to implement for any $\V{y}$ and $A$. }
\item{Necessarily stops (message passing can be unstable and sometimes does not converge, especially in the presence of noise).}
\item{Solutions at finite temperatures $(\{\V{c}_{t} \}_t)$ can be obtained by one iteration of SA. These may be more useful than the optimal solution with minimum distortion, especially in the presence of noise. }
\end{itemize}
%%%%%%%%%%%%%%%%%%%%%%%%
%%%%%%%%%%%%%%%%%%%%%%%%
\paragraph{Disadvantages:}
\begin{itemize}
\item{Emergence of metastable states in the hard parameter region. }
\item{Presence of $\rho$. Tuning this parameter is not trivial: larger values of $\rho$ are better for determining the planted solution and decreasing the level of distortion, but are more likely to yield the metastable state. }
\item{Annealing schedule is arbitrary, and it is not {\it a priori} clear how long the algorithm requires to reach a desired solution. }
\end{itemize}
%%%%%%%%%%%%%%%%%%%%%%%%
These disadvantages may be overcome using a range of techniques. Seeding~\cite{Krzakala:12} is a good candidate for avoiding trapping in the metastable state. What we should do is to change the matrix $A$ to a structured one as described in \cite{Krzakala:12}. Beyond the pair flipping process, we may flip more sparse weights to generate trial moves, which should shorten the time required for efficient sampling. Extended ensemble may also be beneficial. For example, simulating different values of $\rho$ simultaneously enables a wider region of the phase space to be sampled, which may help escaping from the metastable state.  In this way, the second disadvantage will be also diminished. Developing extensions for the present formulation of SA will be helpful to control the sparse approximation problem.
%%%%%%%%%%%%%%%%%%%%%%%%%%%%%%%%%%%%%%%%%%%%%%%%%%%%%%
%%%%%%%%%%%%%%%%%%%%%%%%%%%%%%%%%%%%%%%%%%%%%%%%%%%%%%
%%%%%%%%%%%%%%%%%%%%%%%%%%%%%%%%%%%%%%%%%%%%%%%%%%%%%%
\ack
This work was supported by JSPS KAKENHI Grant Numbers 26870185 (TO) and 25120013 (YK).

%%%%%%%%%%%%%%%%%%%%%%%%%%%%%%%%%%%%%%%%%%%%%%%%%%%%%%
%%%%%%%%%%%%%%%%%%%%%%%%%%%%%%%%%%%%%%%%%%%%%%%%%%%%%%
%%%%%%%%%%%%%%%%%%%%%%%%%%%%%%%%%%%%%%%%%%%%%%%%%%%%%%
%\include{bib-jpc}

\end{document}